\documentclass[12pt]{spieman}  % 12pt font required by SPIE;
\usepackage{amsmath,amsfonts,amssymb}
\usepackage{graphicx}
\usepackage{setspace}
\usepackage{tocloft}
\usepackage{lineno}
\usepackage{wasysym}
\usepackage{upgreek}
\usepackage{ulem}
\usepackage[dvipsnames]{xcolor}
\usepackage[T1]{fontenc}

\newcommand*\micron{$\upmu$m}
\newcommand*\water{H$_2$O}
\newcommand*\methane{CH$_4$}
\newcommand*\ammonia{NH$_3$}

\newcommand{\rev}[1]{#1}

\title{The Science Potential of Characterizing Gas Giant Exoplanets with HWO}

\author[a,*]{Beck Dacus}
\author[a]{Jean-Baptiste Ruffio}
\author[a]{Quinn Konopacky}
\author[b,c,d,e]{Renyu Hu}
\author[f]{Tyler D. Robinson}
\author[g]{Mary Anne Limbach}
\author[h]{Kielan Hoch}
\author[i]{Katelyn Horstman}
\author[j,k]{Bruce Macintosh}
\author[i,b]{Dimitri Mawet}
\author[b]{Michael W. McElwain}
\author[l]{Bertrand Mennesson}
\author[b]{Charley Noecker}
\author[h]{Marshall D. Perrin}
\author[h]{Laurent Pueyo}
\author[m,n]{Dmitry Savransky}
\author[l]{Corey Spohn}
\author[h]{Sarah Steiger}
\author[o]{Connor Vancil}
\author[p]{Ji Wang}
\author[j]{Nicole Wolff}
\author[a]{Shelley Wright}

\affil[a]{Department of Astronomy \& Astrophysics, University of California San Diego, La Jolla, CA, USA}
\affil[b]{Jet Propulsion Laboratory, California Institute of Technology, 4800 Oak Grove Drive, Pasadena, CA 91109, USA}
\affil[c]{Department of Astronomy \& Astrophysics, The Pennsylvania State University, University Park,
PA 16802, USA}
\affil[d]{Center for Exoplanets and Habitable Worlds, The Pennsylvania State University, University Park,
PA 16802, USA}
\affil[e]{Institute for Computational and Data Science, The Pennsylvania State University, University Park,
PA 16802, USA}
\affil[f]{Lunar \& Planetary Laboratory, University of Arizona, Tucson, AZ 85721 USA}
\affil[g]{Department of Astronomy, University of Michigan, Ann Arbor, MI 48109, USA}
\affil[h]{Space Telescope Science Institute, 3700 San Martin Drive, Baltimore, MD 21218, USA}
\affil[i]{Department of Astronomy, California Institute of Technology, Pasadena, CA 91125, USA}
\affil[j]{Department of Astronomy \& Astrophysics, University of California, Santa Cruz, CA95064, USA}
\affil[k]{University of California Observatories, 1156 High Street, Santa Cruz, CA 95064, USA}
\affil[l]{NASA Goddard Space Flight Center, Greenbelt, MD 20771, USA}
\affil[m]{Sibley School of Mechanical and Aerospace Engineering, Cornell University, Ithaca, NY, 14853}
\affil[n]{Carl Sagan Institute, Cornell University, Ithaca, NY 14853, USA}
\affil[o]{Department of Physics, University of California, Santa Barbara, CA 93106, USA}
\affil[p]{Department of Astronomy, The Ohio State University, 4055 McPherson Laboratory, 140 West 18th Avenue, Columbus, OH 43210, USA}

\cftpagenumbersoff{figure}
\cftpagenumbersoff{table} 
\begin{document} 
\maketitle
\textbf{}
%https://www.spiedigitallibrary.org/journals/journal-of-astronomical-telescopes-instruments-and-systems/call-for-papers
% This document shows the required format and appearance of a manuscript prepared for SPIE journals. It is prepared using LaTeX2e with the class file \texttt{spieman.cls}. Please note that the following journals require the use of structured abstracts in manuscript submissions: \textit{Biophotonics Discovery}, \textit{Neurophotonics},  \textit{Journal of Biomedical Optics}, and \textit{Journal of Medical Imaging}. Structured abstracts are encouraged for the \textit{Journal of Micro/Nanolithography, MEMS, and MOEMS}. Guidelines are available on the journal website. Whether structured or single-paragraph, the abstract should be a summary of the paper and not an introduction. Because the abstract may be used in abstracting and indexing databases, it should be self-contained (i.e., no numerical references) and substantive in nature, presenting concisely the objectives, methodology used, results obtained, and their significance. A list of up to six keywords should immediately follow.

\begin{abstract} 
%\noindent The Habitable Worlds Observatory (HWO) will do more than find habitable planets - it will transform our understanding of exoplanet atmospheres and planet formation.
\noindent With the ability to directly image Earth-like exoplanets and search \rev{their atmospheres for biosignatures, the upcoming Habitable Worlds Observatory (HWO) will also collect high signal-to-noise ratio (S/N) reflected-light photometry and spectra of nearby gas giant exoplanets. Such high-quality data would allow investigations into the formation and evolution of low-mass, low-temperature gas giants, and could enable the detection of exomoons around these planets. We} use the EXOSIMS direct imaging mission simulator to model HWO observations of \rev{Jupiter-radius gas giants at Earth-like and Jupiter-like instellations around the 164 stars in the ExEP target list (1.3~pc$\,\leq d \leq\,$23~pc).}
We find that \rev{HWO should be able to achieve S/N$\,\geq\,$5 broadband visible-light detections of gas giants in this instellation range within 5 minutes of integration.}
\rev{At Jupiter-like instellations, 10 hours of R~=~1000 near-IR spectroscopy with HWO should reveal water, methane, and ammonia absorption features in the atmospheres of gas giants orbiting any ExEP star, offering unique insight into the formation conditions and histories of gas giants at a wide range of semi-major axes and masses.}
\rev{HWO optical time-series photometry should exceed 1\% flux precision for gas giants at Jupiter's instellation level at $d\leq$~7 parsecs, and for Earth-instellation gas giants around any ExEP target star.}
\rev{Time-series light curves at this cadence and precision could, over tens of hours, reveal rotation-induced variability comparable to Jupiter's.}
\rev{Eclipses of Mars-sized exomoons may be detectable in high-cadence light curves ($\sim$~5 minute exposures) of Jupiter sized planets in the habitable zones of ExEP stars at~$d\leq$~10 parsecs.}
\rev{For any hypothetical Earth-like exomoons with oxygen-rich atmospheres at $d\leq$~7 parsecs from the Solar System, HWO might be able to detect the spectral signature of molecular oxygen amid the parent planet's photon noise in deep ($\sim$~400 hour integration) spectroscopic HWO observations at R~=~1000.}
\rev{Such moons, if they exist, represent additional habitable worlds that HWO could investigate for biosignatures.}
\end{abstract}

% Include a list of up to six keywords after the abstract

% Include email contact information for corresponding author
{\noindent \footnotesize\textbf{*}Beck Dacus,  \linkable{bdacus@ucsd.edu} }

\begin{spacing}{2}   % use double spacing for rest of manuscript

\section{Introduction}
\label{sect:intro}  % \label{} allows reference to this section

% \textbf{Deadline for submission Sep 1st ($<3$ months)}

% Goal: Extend ExoEarth simulations (\url{https://www.overleaf.com/8816235917zvtssxksdkgg#3bdd5a}) to exoJupiters

% Todo list:
% \begin{itemize}
%     \item Read Nemati+2019
% \end{itemize}

% check if they go beyond the dark hole would they work at 1-5au?
% Jupiter at 5au,10pc, would be 0.5as
% See Figure 2 in Nemati+2020 for coronagraph free contrast -> 2e-4 local flux ratio

% If the goal is to reach a normalized intensity of $<$ a few times $10^{-9}$ (relative to core unobscured flux from star) at $\sim0.4$ arcsec (max sep of Jupiter at 10pc; for 4m mirror at $\lambda=550$nm this is $\sim12\lambda/D$), then according to Mennesson+ 2024 a Hybrid Lyot coronagraph might be able to do it on an off-axis monolithic telescope. For other types this is outside the OWA and the paper doesn't really list NI/contrast beyond IWA and OWA.

% \textbf{New Introduction}

% introduction to HWO and exoEarths/biosignatures
Following the recommendations of the 2020 Decadal Survey \cite{decadal2020}, NASA’s astrophysics division has begun developing the Habitable Worlds Observatory (HWO) \cite{feinberg_habitable_2024}, a UV/optical/IR space telescope with a planned launch in the 2040s.
The goal of this mission is to directly image 25 Earth-sized exoplanets (exo-Earths) in the habitable zones (HZs) of nearby Sun-like stars, and to constrain the abundances of biosignature molecules in their atmospheres. Direct imaging of terrestrial exoplanets is not the only area of astrophysics where HWO will excel; the observatory is predicted to be capable of everything from resolving low-mass main-sequence stars in nearby galaxies \cite{smercina_resolving_2025} to refining estimates of the Hubble constant by measuring the distances of Cepheid variables out to 100 Mpc \cite{anand_envisioning_2025}. Nevertheless, the technical challenges and scientific opportunities involved in detecting molecules in exo-Earth atmospheres drive HWO’s ongoing conceptual design.
Among the planets that have been directly imaged at the time of this writing, 51 Eridani b has the most extreme planet-to-star flux ratio ($\sim10^{-6}$ in the L band) \cite{samland_spectral_2017}, while the smallest angular separation is that of $\beta$ Pictoris c (138 mas) \rev{detected via interferometry}\cite{nowak_direct_2020, lagrange_unveiling_2020}. Meanwhile, HWO will need to push limits of direct imaging in both of these aspects simultaneously, imaging exo-Earths $10^{10}$ times fainter than their stars with angular separations on the order of 100 mas. \rev{The goal of the Nancy Grace Roman Space telescope's coronagraphic instrument is to partially bridge this gap, imaging mature gas giants in reflected light with flux ratios as low as $\sim10^{-9}$\cite{krist_end--end_2023}.}

Spectroscopically characterizing the atmospheres of exo-Earths will place even greater demands on HWO's design tolerances. Studies simulating atmospheric retrievals on synthetic, reflected visible-light spectra of Earth-twins \cite{stark_optimized_2024, latouf_bayesian_2023} have found that, at a spectral resolution R~=~140, the average signal-to-noise ratio (S/N) for \textit{each data point} in the spectrum must be~$\geq5$ for water to be detected. Assuming uncorrelated photon noise, this implies a minimum photometric detection S/N of $\sim$~40 for an exo-Earth. With an expected planet count rate of 30 photons per minute \cite{howe_scientific_2024}, detecting and characterizing 25 exo-Earths in less than 2 years will require HWO's coronagraphic instrument to combine unprecedented starlight suppression \cite{mennesson_current_2024}, optical stability, and low-noise detectors \cite{stark_exoearth_2019}, making it one of the most powerful astrophysical observing platforms in history.

The requirements for directly imaging gas giants in reflected light are not as stringent as for exo-Earths. Where an exo-Earth orbiting 1 AU from a Sun-like star at 10 pc would have a maximum angular separation of 100 mas, a Saturn-twin would have a separation of 1". Moreover, despite receiving only 1\% the instellation of an exo-Earth, a Saturn-twin (even without rings) would still be twice as bright due to its $\sim9R_\oplus$ radius and a geometric albedo of $\sim0.5$ \cite{karkoschka_methane_1998}, compared to Earth's 0.2 \cite{mallama_characterization_2009}. Jupiter, at a separation of 0.5" and an albedo of 0.54 \cite{karkoschka_methane_1998}, would be over 12 times brighter than an exo-Earth. All else being equal, HWO observations of a Jupiter-twin would have S/N between 3 and 12 times higher than an exo-Earth. \rev{These data would also be of higher quality than those obtained from the Nancy Grace Roman space telescope, enabling}
% Such high-S/N data would enable
researchers to test planet formation theories for planets near the peak of the gas giant occurrence rate distribution \cite{johnson_giant_2010,howard_occurrence_2010,pascucci_universal_2018,fulton_california_2021}.

Transit/eclipse spectra of hot Jupiters \cite{madhusudhan_carbon-rich_2011,line_systematic_2014,gandhi_retrieval_2023} and emission spectra of directly imaged planetary-mass companions \cite{konopacky_detection_2013, lavie_helios-retrieval_2017,xuan_are_2024, ruffio_jupiter-like_2026} have allowed researchers to estimate what fraction of a planet's mass was accreted from the protoplanetary disk as solid pebbles/planetesimals, and what fraction was accreted as gas. Pebble/core accretion theories predict that gas giants like those in the Solar System formed first from cores composed of solid refractories and condensed volatiles, enhancing their accretion of solids relative to the dust-to-gas mass fraction of the disk itself ($\sim0.01$\cite{birnstiel_gas-_2010}). Direct collapse/gravitational instability formation theories, on the other hand, predict an accreted solid-to-gas mass ratio largely reflecting the proportions of the disk they formed from. Hydrogen and helium are in gaseous form throughout the disk; whether volatile molecules like CO and \water{} exist in solid or gas phases depends on the local temperature of the disk; and refractories will be in solid phase in all but the hottest, innermost parts of the disk. While past studies have primarily focused on the atmospheric abundances of carbon and oxygen \cite{changeat_five_2022,hoch_assessing_2023, xuan_are_2024}, both of these elements tend to be found in volatile molecules within the disk. Their solid- and gas-phase enrichments will therefore change as a function of distance from the star \cite{oberg_effects_2011,chachan_breaking_2023}, and icy pebbles drifting past these molecules' snow lines may increase \rev{said molecules' local} gas-phase concentration\cite{booth_chemical_2017}. The abundances of carbon and oxygen alone are not enough to constrain the accretion history of giant planets. This makes the abundances of refractory elements, which are much more likely to have accreted as solids, an important aspect of ongoing planet formation studies \cite{schneider_how_2021-1,turrini_tracing_2021}.

The atmospheres of gas giants detectable with HWO are expected to contain ammonia (\ammonia{}) \cite{karkoschka_spectrophotometry_1994, hu_information_2019}, and those at semi-major axes $a_p\sim$1~au will likely contain alkali metals (Na, K) \cite{cahoy_exoplanet_2010}. However, disentangling these trace species will require sufficient spectral resolution to distinguish them from the dominant carbon and oxygen-bearing molecules (e.g., \methane{}, \water{}) at visible and near infrared wavelengths. \rev{M}oderate-to-high resolution (R~$\geq$~1000) \rev{spectroscopy of directly imaged gas giant and brown dwarf thermal emission} has \rev{revealed the detailed composition of their atmospheres} \cite{hoch_assessing_2023, xuan_are_2024, ruffio_jwst-tst_2024}. It can distinguish the unique molecular features in exoplanet atmospheres from the diffracted starlight (speckles) of the host star. For example, at $R\sim2,700$, JWST/NIRSpec recently enabled the first detection of sulfur (H$_2$S) in the atmospheres of the HR~8799 planets \cite{ruffio_jupiter-like_2026}. At the \rev{orbital radii of these planets}, sulfur in the protoplanetary disk would have condensed into a solid; the inferred uniform enrichment of C, O, and S therefore provides direct evidence of efficient solid accretion in massive gas giants ($>5M_{\mathrm{Jup}}$\rev{) on wide orbits (}$>10\,$au). Additionally, $R>1,000$ spectroscopy enables more detailed characterization of exoplanets including radial velocity \cite{ruffio_radial_2019}, spin \cite{wang_detection_2021}, and even isotopologue detections \cite{gandhi_jwst_2023}. [\citenum{ruffio_characterizing_2026}] explores the potential of using moderate-to-high-resolution spectroscopy \rev{with HWO to directly image} exo-Earths and \rev{search for atmospheric} biosignatures. In this work, we expand on the framework outlined in [\citenum{ruffio_characterizing_2026}] to explore the science cases that HWO observations of gas giant exoplanets might enable.

We also explore the possibility of detecting photometric variability caused by gas giant rotation and orbiting exomoons. It has been suggested that, with sufficiently long temporal baselines, high observing cadence, and photometric sensitivity, HWO could use the photometric variability of exo-Earths to constrain their rotation periods \cite{the_luvoir_team_luvoir_2019} and construct longitudinal maps of their surface albedos \cite{knutson_map_2007,cowan_detecting_2025}. [\citenum{crossfield_global_2014}] use Doppler imaging of the Luhman 16 brown dwarf binary at a cadence of 5 minutes to coarsely map the surface of Luhman 16B, and [\citenum{sutlieff_exploring_2024}] find that HD 1160B is variable at the $\sim$~10\% level based on its 18-minute cadence light curve, which the authors attribute to surface rotation. Leveraging HWO's extreme photometric stability, it may be possible to perform similar investigations into the rotation periods and cloud structures of cool gas giants.

[\citenum{limbach_exomoons_2024}] find that HWO may be able to detect exomoons similar in size to Earth's moon orbiting nearby ($\sim$~5 pc) exo-Earth planets by observing 10 instances of the exomoon's disk and/or shadow crossing the face of the planet. The same authors note that contamination from light reflected by exomoons and exorings could potentially alter the appearance of a planet's spectrum, reducing the accuracy of atmospheric retrievals. On the other hand, this necessarily implies that multi-object spectral retrievals may be able to reveal the existence of exomoons directly from their spectral features. While all candidate exomoon signals remain unconfirmed as of this writing \cite{teachey_evidence_2018, kipping_exomoon_2022, unni_doppler_2025}, numerous methods for detecting them have been proposed over the past several decades, including stellar transits \cite{sartoretti_detection_1999}; transit timing/duration variations \cite{kipping_transit_2009}; reflection/emission light curves \cite{limbach_detection_2021} and spectra \cite{williams_looking_2004,oza_redshifted_2024}; radial velocity shifts \cite{Vanderburg2018AJ....156..184V, ruffio_detecting_2023, horstman_rv_2024}; microlensing \cite{sajadian_microlensing_2023}; and astrometry \cite{agol_center_2015, Kral2025arXiv251120091K}. In this work, we focus on (1) photometric variability caused by ``mutual events" like those described in [\citenum{limbach_exomoons_2024}], and (2) on detection from the blended spectra of gas giants and their moons.

%With photometric measurements at high-enough cadence and S/N, it may even be possible to detect the shadows and/or transits of spatially unresolved exomoons \cite{cabrera_detecting_2007, dobos_possibility_2016, forgan_feasibility_nodate, limbach_detection_2021, limbach_exomoons_2024}.

Exomoons represent a new frontier in the study of exoplanetary systems and planet formation.
%The formation of a system of exomoons is directly linked to the formation pathway of the parent planet, so whether exomoon systems around gas giant exoplanets resemble the satellites of Solar System gas giants will depend on how the exoplanetary system formed. The masses and orbital characteristics of moons around gas giants might therefore act as an independent probe of an exoplanet's formation history.
Several moons of Solar System gas giants harbor liquid water oceans beneath thick crusts of ice \cite{reynolds_habitability_1983, nimmo_surfaces_2025}. Furthermore, Saturn’s moon Titan has a thick atmosphere with liquid methane and ethane on its surface \cite{griffith_influx_1995, coustenis_formation_2005, sekine_formation_2013, glein_noble_2015}. The properties of these Solar System moons suggest that terrestrial planets in the HZ may not be the only type of habitable world, and that by studying gas giant exoplanets and their moons, HWO may be able to further its goal of understanding life and habitablility in the cosmos. Given the lack of confirmed exomoons, it is unclear how exomoon properties like mass, orbital radius, and compositon vary with respect to those of their parent planets. This is especially true in the case of planets for which there is no analogue in our Solar System, such as gas giants orbiting in the HZs of their stars (e.g. $\upsilon$ Anromedae d \cite{mcarthur_new_2010} and $\mu$ Arae b \cite{gozdziewski_orbital_2022}). [\citenum{williams_habitable_1997}] find that moons more massive than $0.12M_\oplus$ receiving levels of instellation comparable to Earth could retain nitrogen-rich atmospheres, like those of Earth and Titan, for billions of years. [\citenum{barnes_stability_2002}] find that the orbits of Earth-sized moons around Jovian planets in the HZs of their stars are stable on geologic timescales due to the transfer of angular momentum from the planet's spin to the orbit of the satellite. However, they found this is only true in the HZ of stars more massive than 0.5$M_\odot$, in agreement with [\citenum{heller_exomoon_2012}]. [\citenum{rovira-navarro_tidally_2021}] suggest that, like the early Earth-moon system \cite{heller_habitability_2021}, tidal flexing of an Earth- or Mars-sized exomoon could lead to heat flux at the surface rivaling that of sunlight on Earth, and that it could be sustained for billions of years. This and other sources of heat from the parent planet would lead to the presence of a circumplanetary HZ \cite{dobos_effect_2017,heller_exomoon_2013} that would depend on the joint properties of the host star, parent planet, and exomoon. The fields of exomoon formation and habitability are therefore rich in theoretical predictions but lacking observational data to test them, which HWO observations of gas giants may be able to provide.

%While decades of work has been done on potential exomoon detection methods, including but not limited to stellar transits \cite{sartoretti_detection_1999}, transit timing/duration variations \cite{kipping_transit_2009}, reflection/emission light curves \cite{limbach_detection_2021} and spectra \cite{williams_looking_2004,oza_redshifted_2024}, radial velocity shifts \cite{ruffio_detecting_2023, horstman_rv_2024}, microlensing \cite{sajadian_microlensing_2023}, and astrometry \cite{agol_center_2015}, even the most promising exomoon candidates remain unconfirmed as of this writing. \cite{teachey_evidence_2018, kipping_exomoon_2022, unni_doppler_2025}
 %With HWO's extreme stability and sensitivity, in addition to detecting exomoons comparable to those orbiting Jupiter and Saturn, it may also be able to characterize their surfaces and/or atmospheres. We explore the prospects for exomoon detection with HWO using different techniques and estimate the smallest detectable exomoons as a function of distance from the Sun. 

% Manuscript outline
This manuscript is organized as follows. 
In Section~\ref{sec:methods}, we describe how we use the EXOSIMS direct imaging simulation framework \cite{savransky_wfirst-afta_2015} to model HWO’s high-contrast instrument and simulate observations of nearby gas giants. In Section~\ref{sec:detection}, we present S/N estimates for gas giants as a function of host star brightness, heliocentric distance, and exposure time. In Section~\ref{sec:atmo} we present estimates of molecular detection S/N in gas giant atmospheres at different spectral resolutions. \rev{In Section~\ref{sec:lightcurve}, we present HWO's sensitivity to gas giant temporal flux variation and how it could enable planetary rotation rates to be estimated.} In Section~\ref{sec:exomoons}, we then simulate the photometric detection of exomoons via broadband time-series monitoring, and investigate the prospect of characterizing atmospheres on these exomoons. We discuss the results in Section~\ref{sec:discussion}.  Finally, we present concluding remarks and discuss potential future work in Section~\ref{sec:conclusion}.

\section{Methods} \label{sec:methods}

In this section, we present the framework we use to simulate HWO observations of gas giant exoplanets around nearby stars. We then outline how we use the outputs of these simulations to estimate the signal-to-noise ratio (S/N) for: detection of the planet's broadband emission; absorption from atmospheric molecules; and variability in the flux of the planet over time.

\subsection{Target Stars} \label{sec:exep}
\begin{figure}
\begin{center}
\includegraphics[width=\linewidth]{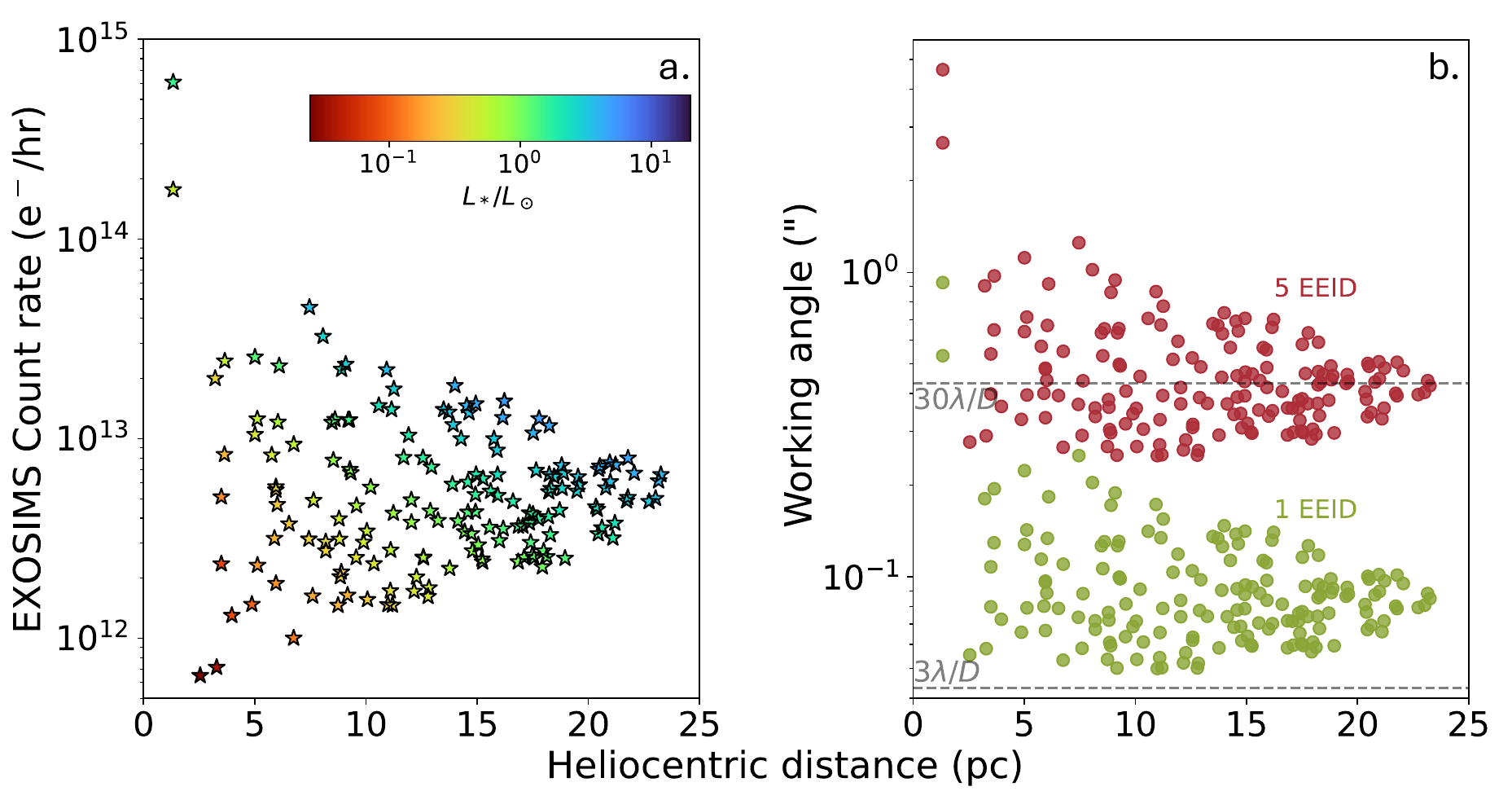}
\end{center}
\caption 
{ \label{fig:exep}
Properties of the ExEP candidate target stars \cite{mamajek_nasa_2024} in our simulation as a function of their distance from the Sun. (a.) Electron count rate from each star \textit{prior to the coronagraph}, assuming $\lambda_c$~=~\rev{550nm} and BW~=~20\% (see Figure~\ref{fig:astrophsources}). The color of each point represents the bolometric luminosity of that star in units of solar luminosities. (b.) Angular separation corresponding to a planet at quadrature \rev{phase} around each star, assuming 1 (green) and 5 (red) Earth-equivalent instellation distances (see Section~\ref{sec:exep}) Horizontal dashed lines indicating 3 and 30 $\lambda/D$ are for a telescope with a 7.87-meter diameter at a wavelength of \rev{550nm} (see Section~\ref{sec:detection}.)
} 
\end{figure}

In each simulation, we place one planet around each of the 164 stars identified by the NASA Exoplanet Exploration Program (ExEP) \cite{mamajek_nasa_2024} as promising targets for HWO's exo-Earth search. These stars are all less than 25 parsecs from the Solar System, \rev{with luminosities ranging from 0.02~$L_{\odot}$ (Lalande 21185, M2V, $d$~=~2.6 pc) to 5.2~$L_{\odot}$ (HR 4102, F3V, $d$~=~16.2 pc).}
%and were chosen such that their
%\rev{luminosities} %effective temperatures
\rev{The average luminosity of these stars increases} with heliocentric distance\rev{ (see Figure~\ref{fig:exep}a), ensuring} that their respective HZs remain largely outside the inner working angle (IWA) of HWO's coronagraphic instrument.

% \begin{figure}
% \begin{center}
% \includegraphics[width=0.5\linewidth]{figs/counts_plot.png}
% \end{center}
% \caption 
% { \label{fig:astrophsources}
% Electron count rate in the 500~-~700nm band for different astrophysical sources. The unobstructed count rates for the ExEP target stars are shown with colors corresponding to their luminosities to show how their intrinsic brightness correlates with distance. The zodiacal light value is plotted with a horizontal dark orange line.
% } 
% \end{figure}

For each simulation, one planet is placed around each star at quadrature phase and maximum separation. The albedo spectra (see Section~\ref{sec:planets}), radii, phase angles ($\beta$), and instellation levels are kept identical for each system. Going forward, we report planet-star separations in units of Earth-equivalent instellation distances:
\begin{equation} \label{eq:eeid}
    \dfrac{a_p}{1\mbox{EEID}} = \dfrac{a_p}{1\mbox{AU}} \sqrt{L_* / L_\odot }.
\end{equation}
where $a_p$ is the semimajor axis of the planet and $ L_* / L_\odot $ is the ratio of host star luminosity to the luminosity of the Sun (see Figure~\ref{fig:exep}). This distance metric scales linearly with physical separation around a single star, while allowing more direct comparison of the physical conditions on simulated planets around different stars. We choose to hold the instellation constant rather than the physical separation because (1) the model albedo spectra were calculated assuming a specific instellation, and (2) this ensures that all the planets in a given simulation have the same reflected light luminosity. \rev{}

\rev{Because we simulate planets at fixed instellation, necessitating that their physical separation increase with host star luminosity (see Section~\ref{sec:planets}), their physical projected separation also increases with distance from the Solar System. This leads to correlations between a target star/planet's distance from the Solar System and other properties of the target sample. For instance, if we assume a fixed planet instellation, the flux ratio between planet and star is inversely proportional to the luminosity of the star. Speckle noise will therefore be more significant for planets orbiting more distant, brighter stars, even if they are all at roughly the same angular separation. More luminous stars will also have higher effective temperatures, emitting more energy at shorter wavelengths and potentially changing the optimal bandpass in which to search for orbiting planets; for simplicity, we assume a fixed observing bandpass within each simulation.}

\rev{The flux from exozodiacal dust (exozodi) colocated with the planet will depend on both the instellation and the local dust density. For each ExEP star, EXOSIMS calculates the color-corrected exozodi surface brightness at 1 AU physical separation\cite{spohn_refined_2025}; we multiply this value by 3 in accordance with the median exozodi level found by [\citenum{ertel_hosts_2020}], then scale it by the inverse square of each simulated planet's semi-major axis (i.e. assume a constant dust surface density) to calculate the exozodi surface brightness at the planet location.}
%\rev{according to the inverse-square law to compute assume a constant dust surface density in each target system, leading to an inverse-square relation between surface brightness and semi-major axis.  find that the median dust surface brightness in the HZs of nearby bright stars is roughly 3 times that of our Solar System; we anchor the exozodi surface brightness at 1 AU from the host star to this value. Anchoring the exozodi surface brightness by physical radius, while placing simulated planets according to instellation,}
% We note that this choice leads to increasing projected separation between the planet and the star with increasing heliocentric distance; because of how we scale the exozodi magnitude with physical separation, this
%introduces a dependence of the exozodi
%\rev{flux with distance from the Solar System.} % magnitude with distance.
%This can impact the relative importance of exozodi and residual starlight noise sources as a function of stellar distance (see Figure~\ref{fig:stddev_astro}). Otherwise, for a fixed stellar type and planet instellation, the photon count rate of the exozodi is independent of distance (see Figure~23 in [\citenum{mennesson_current_2024}]), but \rev{for} future HWO observations it might vary by more than an order of magnitude due to the target list selection.
\rev{For solar system zodiacal light, we assume a surface brightness equal to that measured in the ecliptic plane and assume it does not vary with pointing.}

% \rev{We assume a zodiacal light background that does not vary with pointing. For exozodiacal light, we assume a dust density three times that of the Solar System that is constant throughtout the target star system. The flux from the exozodi in a fixed solid angle therefore scales directly with the incident instellation on that dust, i.e. with the inverse square of physical separation from the star. Rather than simulating the effects of inclined dust disks in our simulations, we adopt the above pessimistic dust surface density for all target systems.}

\subsection{Target Planets} \label{sec:planets}

\begin{figure}
\begin{center}
\includegraphics[width=\linewidth]{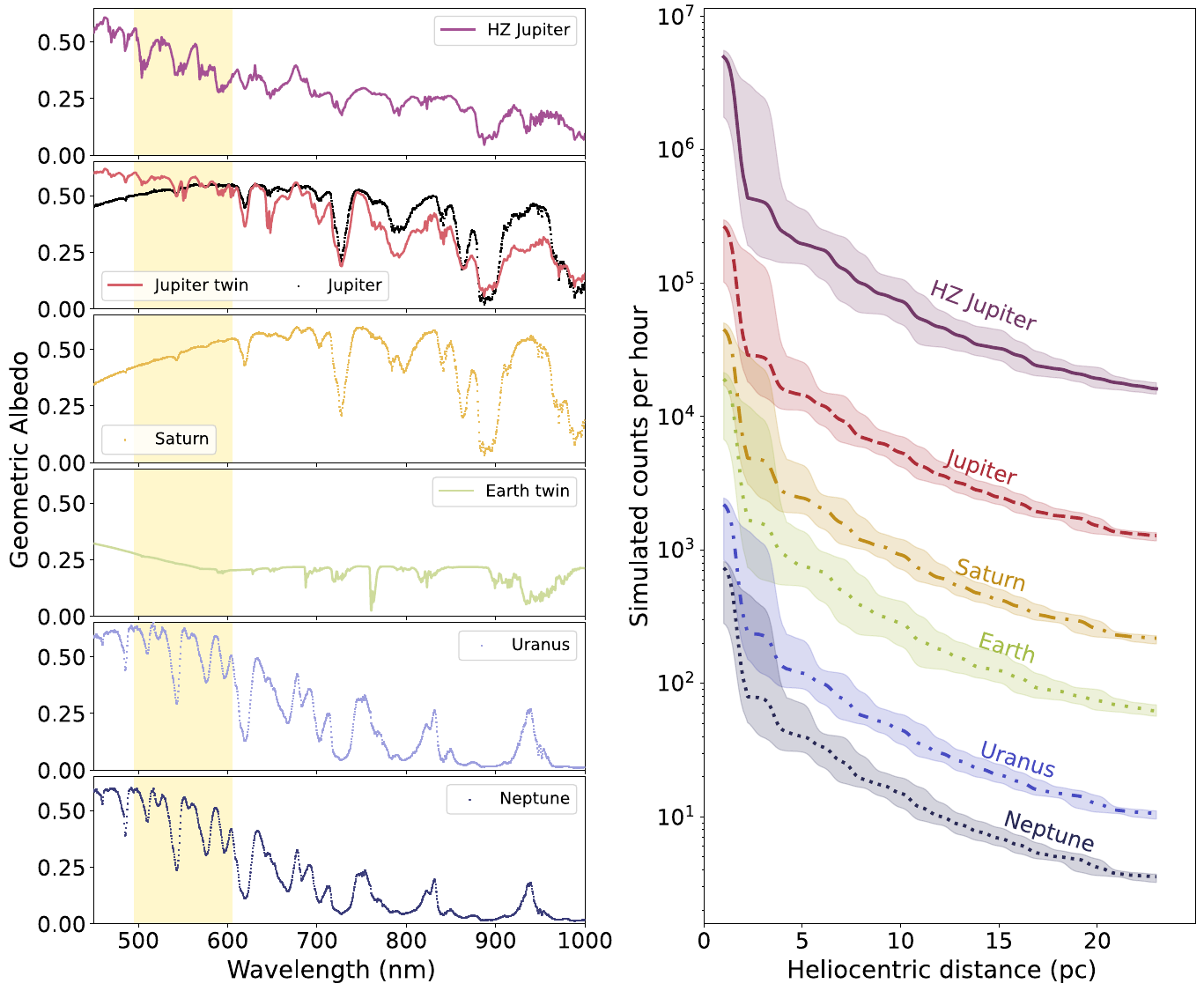}
\end{center}
\caption 
{ \label{fig:astrophsources}
Properties of the planets used in our EXOSIMS simulations. (a.) Geometric albedo of Solar System planets as a function of wavelength. Solid lines are spectral models broadened to R = 1000, while square markers are albedo measurements from \rev{[\citenum{karkoschka_methane_1998}]}, sampled at \rev{0.4}~nm intervals. The golden shaded region is the range of wavelengths from which the count rates in (b.) where calculated. (b.) Electron count rate for Solar System planets as a function of distance from the Sun (includes off-axis coronagraph transmission). Each planet receives the same instellation from its host star as its Solar System counterpart does from the Sun. Lines represent the median value in each distance bin (\rev{4}-pc-wide bins \rev{at 1-parsec intervals, smoothed with a Gaussian filter}), while the shaded regions are bounded by the 16th and 84th percentiles in the same bins.
} 
\end{figure}

In our simulations, we model observations of Jupiter-like planets at Earth-like instellation (``HZ Jupiters") and Jupiter-like instellation (Jupiter twins); these model spectra are shown in the two top-left panels of Figure~\ref{fig:astrophsources}. Each spectral template was generated using the self-consistent atmospheric radiative-transfer and cloud formation model presented in [\citenum{hu_information_2019}]. These models account for the temperature-dependent behavior of water and ammonia condensation, and predict a spectrum for Jupiter that is broadly consistent with measurements of Jupiter's albedo \cite{karkoschka_spectrophotometry_1994, karkoschka_methane_1998} (see Figure~\ref{fig:astrophsources}).

This model uses the molecular line lists of \methane{}, \water{}, \ammonia{}, and relevant collision-induced absorption opacities in the \rev{HITRAN 2020 database \cite{gordon_hitran2020_2022}. For cold/temperate gas giants with low internal heat fluxes, \methane{}, \water{}, and \ammonia{} are expected to be the primary carriers of carbon, oxygen, and nitrogen respectively,\cite{hu_photochemistry_2021} three chemical elements that have been used extensively in planet formation studies (see Section~\ref{sect:intro}). For both HZ Jupiters and Jupiter twins, all three molecular absorbers are assumed to have 3 times solar abundance in the deep atmosphere. The planets' atmospheres are further assumed to be well-mixed, leading to mixing ratios that are largely flat with pressure/altitude but that may jump sharply across cloud decks. For instance, in the HZ Jupiter model atmosphere, \water{} condenses into clouds at 0.1--0.5 bar, depleting to a mixing ratio of $2\cdot10^{-4}$ at higher altitudes, while \ammonia{} does not condense and is therefore present at 3 times solar abundance at all pressures. Meanwhile, in the Jupiter twin model, \ammonia{} condenses into a second cloud deck above water, depleting to a mixing ratio of $5\cdot10^{-7}$ at higher altitudes. Above the ammonia cloud deck, the mixing ratio of \water{} is assumed to be vanishingly small, in agreement with the measurements of Jupiter's atmospheric water found by the Galileo \cite{niemann_composition_1998} and Juno \cite{li_water_2020} spacecraft.}

\rev{The model spectra have a native resolution of R~$\sim$~150,000 across the visible and near-IR, with the exception of \methane{}; at wavelengths shorter than 750nm, we use the \methane{} opacities from [\citenum{karkoschka_methane_2010}] with R~$\sim$~1,000.} Broadband template-matching S/N is most comparable to photometric imaging S/N at low spectral resolution (see Section~\ref{sec:snr_bb}), so \rev{we use broadband template matching S/N at low spectral resolution (R~=~100) as a proxy for the S/N that a similar high-contrast imaging instrument would obtain}. We use the absorption spectra of each individual molecule when calculating its detection S/N (see Section~\ref{sec:snr_mol}).

\begin{figure}
\begin{center}
\includegraphics[width=\linewidth]{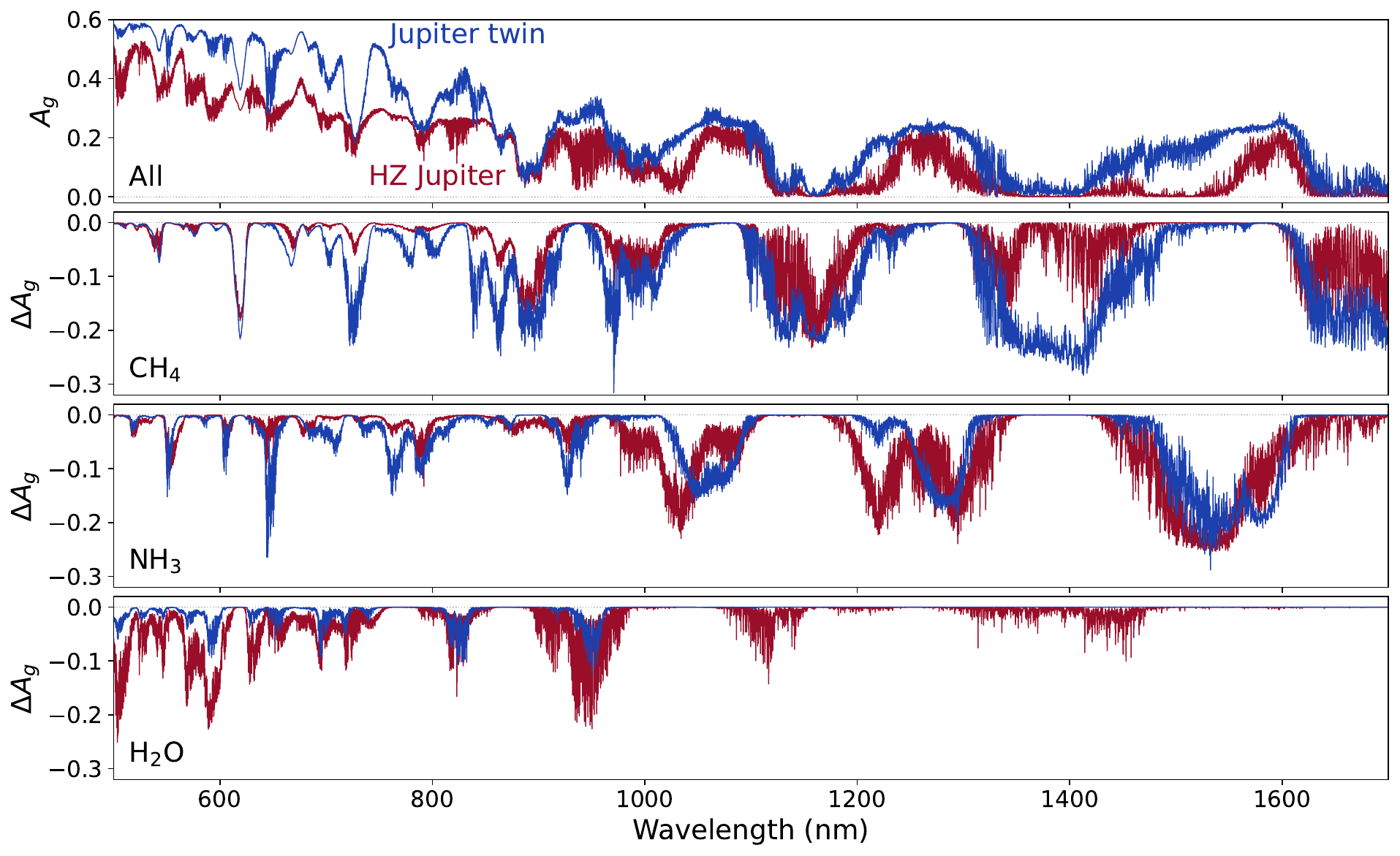}
\end{center}
\caption 
{ \label{fig:mol_templates}
\rev{Template albedo spectra used when simulating observations of HZ Jupiters (red) and a Jupiter twins (blue), convolved to a spectral resolution of 10,000. Top panel: full albedo spectra with cloud reflection, Rayleigh scattering, and molecular absorption included. Bottom three panels: Albedo changes in each spectrum caused by the presence of methane, ammonia, and water, respectively. In each panel, zero albedo or albedo change is denoted with a horizontal dotted line.}}
\end{figure}

Figure~\ref{fig:astrophsources} shows the template albedo spectrum for an HZ Jupiter and Jupiter twin, broadened to a resolving power of 1000, with Jupiter albedo measurements from [\citenum{karkoschka_spectrophotometry_1994}] overplotted in the \rev{0.45} - 1 $\mu$m region. Figure \ref{fig:mol_templates} shows \rev{the HZ Jupiter and Jupiter twin templates over a wider range of wavelengths and higher spectral resolution (R~=~10,000), as well as} the contributions of individual \rev{molecular absorbers} (\methane{}, \water{}, and \ammonia{}) to the total albedo spectrum. Compared to the Jupiter twin, the spectrum of the HZ Jupiter planet \rev{tends to have an overall lower albedo and exhibit} more pronounced molecular absorption bands. The increased temperature of HZ Jupiter atmospheres causes ammonia that otherwise would have condensed into clouds to remain in vapor form, simultaneously removing a high-albedo cloud deck and increasing the column density of absorbing molecular species.

% \begin{figure}
% \begin{center}
% \includegraphics[width=\linewidth]{figs/jupiter_twin_template.pdf}
% \end{center}
% \caption 
% { \label{fig:jupiter_twin_template}
% Model albedo spectra of a Jupiter-analog gas giant exoplanet, broadened to a spectral resolution of 10,000. The top row shows the full atmospheric model spectrum, with measured albedos from [\citenum{karkoschka_spectrophotometry_1994}] (R = 500 - 1000) overplotted in orange for reference. The bottom three rows display the effects of absorption from specific molecules. The ``envelop" of each spectrum is shown as a dashed gray line. \textcolor{red}{TODO: Label 800 and 1200nm cloud features, and label the model native spectral resolution at different wavelengths. Zoom in on 650nm ammonia feature and nearby water features and show different spectral resolutions.}
% } 
% \end{figure}

% \begin{figure}
% \begin{center}
% \includegraphics[width=\linewidth]{figs/hz_jovian_template.pdf}
% \end{center}
% \caption 
% { \label{fig:hz_jovian_template}
% Same as Figure~\ref{fig:jupiter_twin_template} but for a simulated HZ Jupiter.
% } 
% \end{figure}

\subsection{Simulating HWO} \label{sec:exosims}
 
% In preparation for future yield studies, we compute HWO's completeness in detecting gas giant exoplanets, detecting molecules in their atmospheres, and identifying intrinsic variability in their light curves. We do this by simulating observations of each type of planet described in Section~\ref{sec:planets}, placed $\beta=\pi/2$ at their respective insolation distances.

To simulate HWO observations of the modeled targets, we employ the EXOSIMS \rev{direct imaging mission simulator}. \cite{savransky_wfirst-afta_2015, delacroix_science_2016} 
% Specifically, we make use of the ``MHRS" (moderate-to-high-resolution spectroscopy) branch on GitHub. 
For a detailed description of the \rev{moderate-to-high-resolution spectroscopy (MHRS)} implementation of EXOSIMS, we refer the reader to [\citenum{ruffio_characterizing_2026}]; we briefly summarize here.

In each simulation, we use the principles outlined in Section~\ref{sec:planets} to specify the projected angular separation of each simulated planet. This angular separation determines the magnitude of the local starlight, local coronagraphic starlight suppression, and local surface brightness of the exozodi. All counts are calculated for a 2x2-pixel aperture with an on-sky solid angle of $2.26(\lambda_c/D_{eff})^2$, where $\lambda_c$ is the central wavelength of the observing bandpass and $D_{eff}$ is the effective diameter of the telescope primary mirror.

The speckle contribution to the aperture counts are calculated assuming a charge-6 vector vortex coronagraph as presented in [\citenum{belikov_coronagraph_2024, mawet_apodized_2025}], with a contrast floor of $10^{-10}$ in alignment with previous studies \cite{stark_paths_2024}. The coronagraph is assumed to have a transmission of 0.8 for extended sources, while the transmission for off-axis point sources (i.e. planets) is assumed to rise with working angle until reaching a maximum of 0.48.
Because research into optimizing the inner and outer working angles (IWA and OWA, respectively) of HWO’s coronagraph is ongoing, we do not set an IWA or OWA in our simulations. At small angular separations, we instead use the variable off-axis transmission function of the coronagraph as a ``soft” IWA, allowing us to explore the tradeoff between planet brightness and coronagraphic throughput.

When detecting and characterizing Jupiter-like exoplanets, the OWA of HWO’s coronagraph will have more influence on the kinds of planets it can detect than the apparent magnitude of the planets themselves. For instance, at quadrature phase, Saturn (without its rings) is twice as bright as Earth in visible reflected light. However, with an OWA of 0.85", a planet 10 pc from the Solar System orbiting its star at 10 au would spend at most 14\% of its orbit both within the dark hole and with its observer-pointing face illuminated; this fraction only goes down for planets orbiting stars nearer to the Sun. Such planets also orbit their stars much more slowly than planets in the HZ, so multi-epoch observations may not significantly increase the chances of detecting them. We chose not to place an OWA limit in this study to show the achievable S/N values for gas giants orbiting at angular separations $\gtrsim$~30~$\lambda/D$ (see Figure~\ref{fig:exep}).

%. To quickly summarize, Ruffio et al. (in review) makes use of a S/N definition based on template-matching for both planet and molecular detection S/N.The broadband detection S/N of an object from template matching is typically not strongly dependent on spectral resolution as long as the detector noise introduced by adding pixels is negligible compared to other sources. This is because the fine spectral features at high spectral resolution do not represent a large fraction of the planet photons when compared to the continuum. Because of this, we do not separately simulate a high-contrast imaging instrument. 

% When modeling HWO’s coronagraph instrument, we assume the use of an integral field spectrograph (IFS). Many HWO design studies assume the use of a high-contrast imager for initial exo-Earth detection, with follow-up characterization performed by a slit- or fiber-fed spectrograph [cite]. We once again refer the reader to Ruffio et al. (in review) for an explanation of why an IFS may in fact be a viable choice for HWO’s primary mission instrument, and why at low spectral resolution (R$\lesssim1000$) and/or negligible detector noise, the template-matching detection S/N should closely correspond to the of an imager. We refer the reader to Ruffio et al. (in review) (Table~2 and 3 therein) for details about the properties we assume for HWO and its scientific instrument, but review our fundamental assumptions here.

We assume a circumscribed diameter of 7.87 meters when modeling HWO’s primary mirror, as is used in [\citenum{belikov_coronagraph_2024}], resulting in a spatial resolution of 12 milliarcseconds at $\lambda=$~550 nm. We multiply the collecting area of this mirror by an ``obscuration factor” of 0.88 when calculating the total flux received from each source to encapsulate the effect of possible mirror segmentation and/or obscuration by support structures. 
%While the outer working angle (OWA) of the dark hole does not impact habitable zone planets, it sets a limit on the outer most planets that can be detected around the closest stars to the Sun. In this study, we set the OWA to be 10 arcseconds, larger than the largest angular separation of the planets in our simulation. This is motivated by the desire to explore the full range of possible planets in the solar neighborhood in this study to inform science cases, even though the OWA could be smaller in practice.  We source coronagraphic on- and off-axis throughputs from the charge-six optical vortex coronagraph parameters given in [\citenum{belikov_coronagraph_2024}].   While the off-axis core throughput in these files drops to 0 at separations $\gtrsim30\lambda/D$, we assume a constant off-axis core throughput of 0.48 beyond this limit. We then fix the level of starlight suppression at the planet location to either $10^{-9}$ or $10^{-10}$.

\subsubsection{S/N Per Spectral Bin} \label{sec:snr_bin}

Because we are assuming that \rev{the mean count rates from the planet and the persistent noise sources are modeled accurately (excepting the speckles), we define ``signal" to mean the number of counts originating from the planet (or equivalently the flux deficit caused by a molecular absorber or temporal variation), and assume the uncorrelated noise in each bin is Poisson-distributed.} \rev{A pixel with index $i$ that receives counts at a rate $C_i$ will then have a corresponding uncorrelated noise variance of $\sigma_{ii}^2=C_i t_{exp}$} (the $ii$ subscript denotes uncorrelated noise). \rev{If the mean count rate from the planet is $P_i$, the} signal-to-noise ratio in that pixel is therefore 
\begin{equation} \label{eq:snr_bin}
    \mbox{S/N}_i = \rev{\dfrac{P_i\sqrt{t_{exp}}}{\sqrt{C_i}}}.
\end{equation}

We refer to $\mbox{S/N}_i$ as the ``S/N per resolving element.” When applied to one spectral resolving element, $\mbox{S/N}_i$ can provide an estimate of the minimum depth of a detectable absorption line. However, looking at individual channels in a spectrum does not leverage the full extent of the information that a spectrum contains with the complete shape of the molecular spectral features; Section~\ref{sec:snr_bb} goes into more depth on how to fold this additional information into the definition of S/N.

\subsubsection{Broadband S/N} \label{sec:snr_bb}

Template matching provides a consistent framework, based on maximum-likelihood, to compute the detection S/N of a planet or even specific molecules \cite{ruffio_characterizing_2026} (see Section~\ref{sec:snr_mol}) \rev{from a spectrum and its errors. It accounts for the fact that, while the counts from photon noise in a single spectral channel are independent from the noise counts in every other spectral channel, counts originating from a planet would be correlated across the entire wavelength range of the spectrum. Therefore, in a spectrum with a large number of wavelength samples, random photon noise is unlikely to mimic the spectral signature of a planet or molecule, even if it could explain the flux residuals within each spectral channel individually. The data points in a spectrum must then be treated as an ensemble rather than in isolation to fully characterize the strength of a planetary signal. This is especially important when considering planetary and molecular detection/characterization at different spectral resolutions; while increased spectral resolution decreases the S/N in individual spectral bins, it can also provide additional information that can be useful for atmospheric model fitting. Such model fits will be key for HWO to place meaningful constraints on the abundances of molecules in exoplanetary atmospheres, especially biosignatures in the atmospheres of exo-Earths.}

\rev{That being said, not all noise counts in a spectrum will be pure Poisson noise, and will therefore not be fully independent from the noise counts in all other spectral channels. For instance, HWO's extremely stable point-spread function (PSF) should allow for the majority of the speckles overlapping with the planet's location to be subtracted; however, imperfect PSF modeling will cause residuals from the star's chromatic PSF remain as smooth flux oscillations in the spectrum, with systematic noise counts in adjacent wavelength samples being correlated with each other. While there can be other sources of correlated noise, we choose to only to include the effects of speckle residuals for simplicity, encapsulating their distribution with the covariance matrix $\boldsymbol{\Sigma}$ (see Section~\ref{sec:spectrum}). In the worst-case scenario, the ``length" scale of the correlation in wavelength space is similar to the FWHM of the planet's spectral features, at which point correlated noise can mimic genuine spectral lines. This highlights the importance of MHRS, as decreasing the width of an instrument's line spread function below the correlation scale can prevent correlated noise from mimicking narrow absorption features.}

In this formalism, we represent the planet spectrum as a vector $\mathbf{p}=[p_1, p_2...p_k]^\top$ where $p_i$ is the flux of the planet in a spectral bin. We assume that all systematic noise sources have been subtracted, so the noise is assumed to be Gaussian and represented by a covariance matrix $\boldsymbol{\Sigma}$.
% (see Section~\ref{sec:spectrum}).
It can then be shown via maximum likelihood estimation \cite{ruffio_characterizing_2026} that S/N of the planet's broadband detection
\begin{equation} \label{eq:snr_tm}
    \mbox{S/N}_{bb} = \sqrt{\mathbf{p} \boldsymbol{\Sigma}^{-1} \mathbf{p}}.
\end{equation}
We note that the broadband template matching S/N for planet detection converges to the imaging S/N at low spectral resolution. We therefore treat S/N$_{bb}$ values at low R as being nearly equivalent to broadband photometric S/N values for our purposes.

\rev{In each of our simulations, we set a constant integration time for every target and calculated the resulting S/N. To calculate the amount of time that a gas giant observing campaign would require, we would need to invert the problem, setting a target S/N and from there identifying the necessary exposure time for each target. For uncorrelated Poisson noise this problem is straightforward, but the presence of correlated noise makes the template matching S/N a non-trivial function of integration time. We would therefore need to find the integration time by iteratively calculating the S/N for multiple integration times, significantly increasing the computation time required for each observing simulation. We therefore use a uniform integration time in each simulation to reduce complexity and computation time, but we plan to implement this functionality into EXOSIMS in the future.}

\subsubsection{Molecular Absorption S/N} \label{sec:snr_mol}

When solely trying to detect the presence of molecular absorption features, we are not interested in the overall continuum flux of the planet, but only in the flux deficit caused by an individual molecule. At each wavelength, this is proportional to the amount by which the presence of a molecule changes the albedo of the model spectrum, in other words $|A_p-A_{no\,mol}|$. \rev{We generate molecular template spectra by subtracting ``leave-one-out" albedo spectra-- in which either \methane{}, \ammonia{}, or \water{} has been left out of the radiative transfer model-- from our primary model spectra where all three molecules are included. This gives us the wavelength-dependent albedo change caused by an individual molecule, which we will call $\Delta A_{mol}(\lambda)$.} By sampling \rev{$\Delta A_{mol}(\lambda)$} at the wavelengths of the instrument we then obtain the vector $\mathbf{p}_{mol}$, which we can use to estimate the molecular detection S/N in a manner analogous to Equation~\ref{eq:snr_tm}:
\begin{equation} \label{eq:snr_mol}
    \mbox{S/N}_{mol} = \sqrt{\mathbf{p}_{mol} \boldsymbol{\Sigma}^{-1} \mathbf{p}_{mol}}.
\end{equation}

\subsubsection{Photometric Variability} \label{sec:photvar}

The S/N of photometric variability over time can be defined with the same formalism as the S/N of molecular absorption (see Section \ref{sec:snr_mol}). In this case, the ``continuum” is analogous to the mean flux of the planet (and its rings/moons if present), and the signal is the alteration to that expected flux caused by the planet’s rotation, changes in its cloud structure, and/or eclipses and transits between a planet and a moon. When calculating the S/N of one type of photometric variability, other sources of variability that have not been subtracted can act like correlated noise that reduces the S/N of the variability interest; however, defining a covariance matrix for such variability is beyond the scope of this paper, so we focus on exploring the S/N of each type of variability separately.

We once again assume that the flux of all persistent noise sources--- which in this case includes the average flux from the planet--- has been subtracted, leaving the vector $\mathbf{F}=[F_1,F_2…F_K]$, where $F_j$ is a measurement of the planet’s flux at time index $j$ and $K$ is the number of exposures in the light curve. If the extent of the variability is small compared to the average flux of the planet, then the variance is essentially the same for each observation. The S/N of the light curve variability is therefore
\begin{equation} \label{eq:snr_light}
\mbox{S/N}_{lc} = \sqrt{\dfrac{\mathbf{F}^\top\mathbf{F}}{\sigma_{all}^2}}
\end{equation}
where $\sigma_{all}^2$ is the per-exposure variance. We can normalize both $\mathbf{F}$ and $\sigma_{all}^2$ to the mean number of counts, $\sum_ip_i$, yielding $\mathbf{f}$ and $(\mbox{S/N}_{bb})^{-2}$ respectively. Equation~\ref{eq:snr_light} then becomes
\begin{equation} \label{eq:photo_strength}
    \mbox{S/N}_{lc} = \sqrt{\mathbf{f}^\top\mathbf{f}} \cdot \mbox{S/N}_{bb} = H\cdot\mbox{S/N}_{bb}
\end{equation}
where we have defined the ``strength" $H=\sqrt{\mathbf{f}^\top\mathbf{f}}$. For a given S/N threshold for detecting photometric variability $T$, the minimum S/N$_{bb}$ required to achieve it is simply $T/H$.

\subsubsection{Simulating Spectrophotometric Observations} \label{sec:spectrum}

When simulating spectra, we broaden the model albedo spectra in Section~\ref{sec:planets} to the assumed resolution of the instrument, R, before sampling the broadened spectrum at a set of wavelengths $[\lambda_1, \lambda_2...\lambda_k]$. The bandwidth BW~$=\dfrac{\lambda_k - \lambda_1}{\lambda_c}$ sets the range of wavelengths in the spectrum and the number of samples $k=$2R$\cdot$BW (the factor of 2 ensures Nyquist sampling). \rev{We then broaden the EXOSIMS template spectrum of the host star to the same resolution, resample it on the same wavelength grid, and multiply the two together.} We represent the result as a vector $\mathbf{p}=[p_1,p_2...p_k]$,
\rev{scaled} % normalized
such that
\begin{equation} \label{eq:norm}
    \sum_i p_i = C_p,
\end{equation}
where $C_p$ is the total electron counts for the planet calculated in Section~\ref{sec:exosims}. \rev{We note that the stellar template spectra in EXOSIMS only have a spectral resolution of $\sim$~500, much lower than the native resolution of our model albedo spectra. For our broadband detection S/N (see Section~\ref{sec:snr_bb}) this is not an issue, since both the stellar template and planet albedo spectrum are broadened to R~=~100. However, when exploring molecular detection S/N (see Section~\ref{sec:snr_mol}) at higher R, we cannot include the effects of narrow stellar absorption lines on the spectrum. In the future we will seek to add higher-resolution stellar template spectra to EXOSIMS to make simulated observations more accurate, and to investigate how HWO could use MHRS to leverage kinematic differences (i.e. radial velocity, spin, etc.) between planets and their stars for detection and characterization.}

% \rev{However, these stellar spectra are of much lower spectral resolution than our model albedo spectra; deep stellar spectral lines are therefore broadened into broader, shallower features. If we were to multiply the planetary albedo spectra by the stellar spectra before scaling them into data units, this would likely have unrealistic effects on both the mean S/N per resolving element (Section~\ref{sec:snr_bin}) and template-matching S/N (Sections~\ref{sec:snr_bb} and~\ref{sec:snr_mol}). The spectral continua of the target stars also do not have large slopes in the 20\% bandpasses used in this work; we therefore do not multiply the planetary albedo spectrum by the stellar spectrum in our simulations. Higher resolution spectra of the ExEP target stars would enable future studies that can account for this effect in simulated observations.}

Counts from background sources and detector noise are calculated similarly \rev{to those of the planet}, using the inbuilt EXOSIMS spectra of (exo)zodiacal dust and the ExEP target stars. Detector noise counts are assumed to be the same at all wavelengths. The combined spectrum\rev{, without Poisson noise,} is therefore
\begin{equation} \label{eq:specvec}
    \mathbf{s} = \mathbf{p} + \mathbf{h} + \mathbf{z} + \mathbf{e}  + \boldsymbol{\xi}
\end{equation}
where $\mathbf{h}$, $\mathbf{z}$, $\mathbf{e}$, and $\boldsymbol{\xi}$ are the sampled spectra of the host star, zodiacal light, exozodi, and detector noise, respectively. Assuming the statistical noise is Poisson-distributed, the variance of the counts at index $i$ is given by $\sigma_i^2 = s_i$.
These values form the diagonal of the covariance matrix $\boldsymbol{\Sigma}$; correlated noise from residual speckles determine the values of the off-diagonal elements of $\boldsymbol{\Sigma}$ (see [\citenum{ruffio_characterizing_2026}] for more details).

\subsection{Exomoon Detection and Characterization} \label{sec:exomoon_methods}

Numerous exomoon detection methods have been suggested in the literature (see Section~\ref{sect:intro}), including those geared specifically toward directly imaged exoplanets  \cite{williams_looking_2004,heller_transits_2016,ruffio_detecting_2023,van_woerkom_spectroastrometric_2024}. Studies concerning HWO in particular have investigated the potential of using time-series light curves to detect exomoons transiting in front of and behind their planets from our perspective \cite{limbach_exomoons_2024}, as well as eclipses where the moon enters the shadow of the planet \cite{limbach_exomoons_2026}. It has also been suggested that HWO could detect exomoons directly through the effect they have on the spectra of their host planets \cite{limbach_exomoons_2024}. In this work, we compare HWO's ability to detect the photometric effects of different transit and eclipse events, as well as direct spectroscopic detection and characterization of large, Earth-like exomoons.

We simulate photometric variability caused by exomoons using a custom code. This software simulates the planet, moon, host star, and observer as a three-dimensional system, and includes the effects of transits as well as eclipses. We assume the existence of only one moon on a circular orbit around the planet, and the planet is assumed to be identical to Jupiter in radius and mass.

\begin{figure}
\begin{center}
\includegraphics[width=0.65\linewidth]{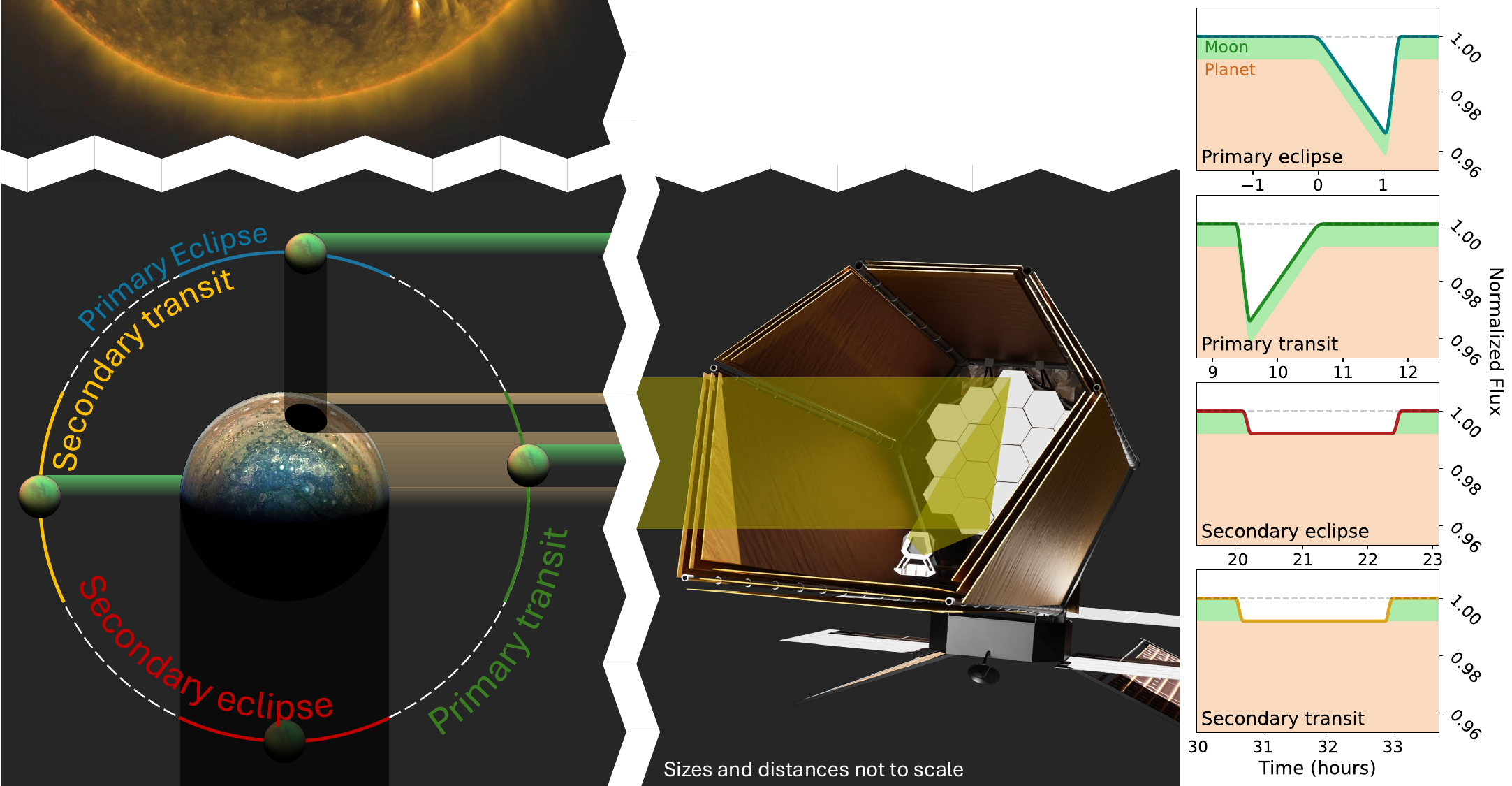}
\end{center}
\caption 
{ \label{fig:mep}
Graphical representation of the four moon-planet mutual events that could be used to search for exomoons (sizes and distances not to scale). \rev{Left and center}: cartoon of the star-planet-moon-observer system, where for clarity all four components are assumed to be in the same plane (this is not a necessary assumption in our simulations). The jagged lines represent large distances separating members of the system. Labels and arcs in different colors highlight at what point in the moon's orbit each event occurs. \rev{Right}: simulated light curves of each event, plotted in the corresponding colors. The moon is assumed to be the size of Earth, orbiting a Jupiter-sized planet with an orbital period equivalent to Io's. The beige shaded region corresponds to the starlight reflected from the planet while the green shaded region represents the light reflected by the moon. (Image credit for Sun/Jupiter/Titan: NASA/SDO; NASA/JPL-Caltech/SwRI/MSSS/Gerald Eichstädt; ESA)
} 
\end{figure}

Previous work has identified four distinct types of alignment, or ``syzygys," which would lead an observer to detect a drop in the combined flux of a planet-moon system \cite{cabrera_detecting_2007,cabrera_detecting_2007-1,berzosa_molina_traces_2018,limbach_exomoons_2024}.The terminology used to refer to these events has varied from publication to publication, and does not always match what is used by the transiting planet community. We therefore choose to employ our own terminology for these four events:
\begin{enumerate}

\item \textbf{Primary eclipse}: when the \textbf{star, exomoon, and exoplanet} are aligned, and the \textbf{exomoon} casts a shadow on the \textbf{exoplanet’s} illuminated face.

\item \textbf{Primary transit}: when the \textbf{observer, exomoon, and exoplanet} are aligned, and the \textbf{exomoon} obscures a portion of the \textbf{exoplanet's} disk from the observer's perspective.

\item \textbf{Secondary eclipse}: when the \textbf{star, exoplanet, and exomoon} are aligned, and the \textbf{exomoon} enters the shadow of the \textbf{exoplanet}, leading it to become completely dark in most cases.

\item \textbf{Secondary transit}: when the \textbf{observer, exoplanet, and exomoon} are aligned, and the \textbf{exoplanet} obscures the disk of the \textbf{exomoon} from the observer's perspective.

\end{enumerate}
Figure~\ref{fig:mep} displays a graphical representation of each mutual event’s geometry.

% \begin{figure}
% \begin{center}
% \includegraphics[width=0.7\linewidth]{figs/mutual_events.pdf}
% \end{center}
% \caption 
% { \label{fig:astrophsources}
% Examples of planet-moon mutual events from results of our custom simulation. TODO: either move to later or take event name labels out
% } 

For moons orbiting an HZ Jupiter, we assume the same albedo that we assume for Earth ($A_{g}=$0.2), since moons at such distances are unlikely to have icy surfaces. This is about twice that of the Moon/Mercury ($A_{g}=$0.12, 0.14), slightly higher than Mars ($A_{g}=$0.17), and much lower than both Io ($A_{g}=$0.63) and Venus ($A_{g}=$0.75). In addition to being representative of an Earth-like moon, $A_{g}=$0.2 represents a middle ground among the albedos of ice-free terrestrial worlds in our Solar System.

\section{Gas Giant Detection} \label{sec:detection}

In this section, we detail the detection S/N that HWO can achieve for Jupiter-like exoplanets in broadband light.
% The purpose of this not to show how easy it is to detect Jovians, but to show which stars a given type of Jovian planet will be detectable around. Jovian planets at instellations comparable to an exo-Earth will face the same IWA limitations, and widely separated Jupiters may fall outside of the OWA; certain Jovians may also fall below the observatory’s contrast limit. All these factors will determine the nearest and furthest distances at which these different Jovians can be detected, which we choose as our figure of merit.
For our bandpass we assume a central wavelength $\lambda_c$ = \rev{550nm} and a bandwidth $\mbox{BW} = 20\%$. The albedos of Jupiter and Saturn \rev{are both high in} this wavelength range (see [\citenum{karkoschka_spectrophotometry_1994}] and Figure~\ref{fig:astrophsources}). \rev{While their albedos peak closer to 700nm, the telescope's point-spread function (PSF), and by extension its inner working angle (IWA), would be roughly 25\% wider at these wavelengths. The PSF of the planet would be wider by the same amount, forcing the size of the photometric aperture to increase, which would then capture more light from diffuse noise sources like Solar System zodiacal light and exozodi. We therefore assume that shorter wavelengths will be favored for initial exoplanet detection.}

\rev{In this section we explore starlight suppression floors of $\zeta_{floor}=10^{-9}$ and $10^{-10}$, bracketing the flux ratio of an Earth-like planet orbiting a Sun-like star ($\sim3\cdot10^{-10}$). We further test PSF stability limits of 10\% (gain $g=10^{-1}$) and 1\% ($g=10^-2$). [\citenum{ruffio_jwst-tst_2024}] found that the wavefront error in their ~2.9~--~5.3\micron{} JWST NIRSpec observations did not exceed 7nm; we assume that HWO's wavefront error in optical wavelengths will allow for similar or greater PSF stability in high-contrast imaging observations. We set a constant integration time $t_{int}$~=~10 hours for each planet, corresponding to the amount of time HWO might spend on a single visit when searching for exo-Earths.}

% We assume a starlight suppression factor of $10^{-10}$, a post-processing gain of $10^{-2}$, and a spectral resolution of 100. We test integration times of $t_{int}$ = 10 hours (representing one visit as part of the exo-Earth search) and 400 hours (representing an exo-Earth deep atmospheric characterization and biosignature searches.)

% \subsection{Noise Sources} \label{sec:noise}

\begin{figure}
\begin{center}
\includegraphics[width=\linewidth]{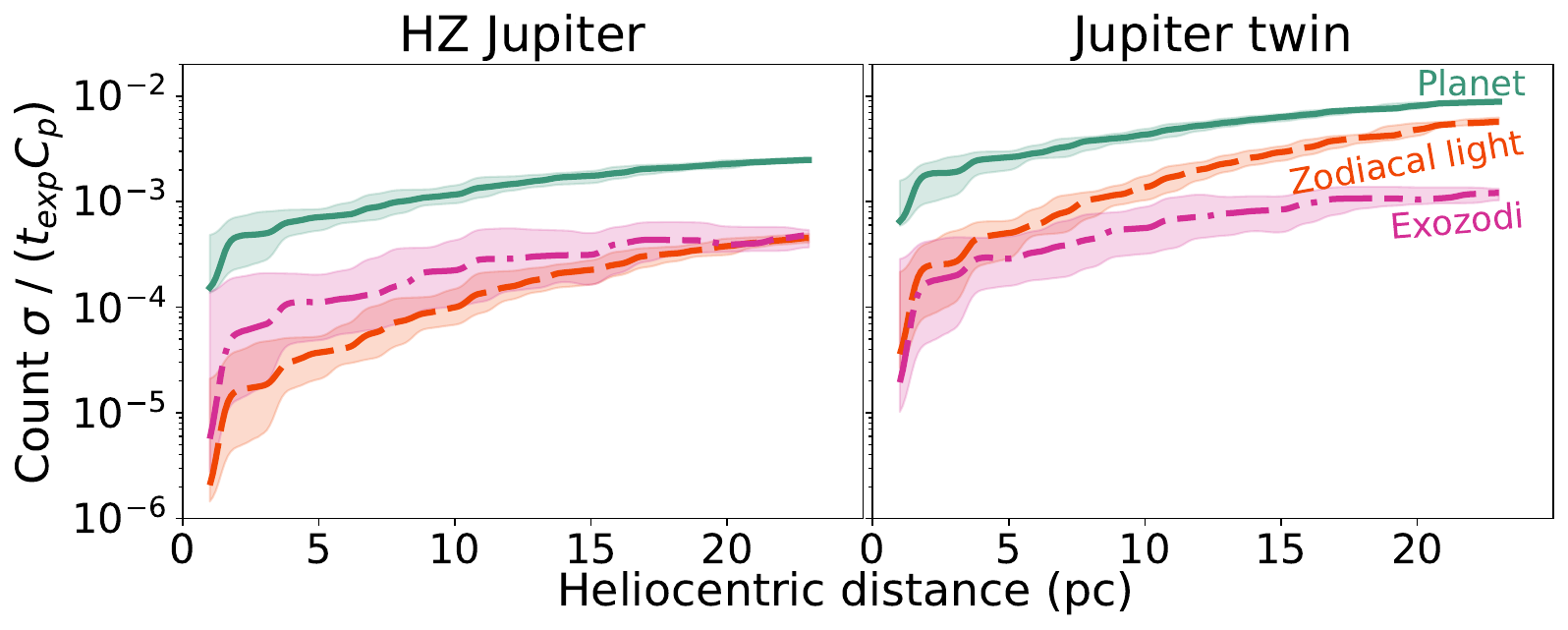}
\end{center}
\caption 
{ \label{fig:stddev_astro}
\rev{Noise counts from astrophysical sources, normalized to the total counts from each gas giant planet, in simulated HWO observations as a function of heliocentric distance. The integration time used in these simulations was 10 hours, with a central wavelength of 550nm ad a bandpass width of 20\%. The standard deviation of the counts from each source is Poisson distributed, i.e. the square root of the total number of counts from each source. Data are binned in 1 pc increments with a bin width of 4 pc, then smoothed with a Gaussian filter. Colored lines trace the median values of each bin, and shaded regions are bounded by the 16th and 84th percentiles.}
} 
\end{figure}

\rev{Figure~\ref{fig:stddev_astro} shows the statistical noise contributions from simulated astrophysical sources, normalized to the total counts from each simulated target planet. The values in the numerator correspond to Poisson noise, hence the line labeled ``Planet" is essentially showing $(t_{int}C_p)^{-1/2}$. All else being equal, the flux from planets and exozodi alike scale with the inverse square of their physical separation from their host star (see Section~\ref{sec:planets}); this is why, for both simulated HZ Jupiters and Jupiter twins, the median Poisson noise contribution from the planets themselves are roughly an order of magnitude greater than their local exozodi. On the other hand, the count rate from zodiacal light in our simulations is constant for all targets; this noise source therefore becomes significant for fainter planets, i.e. with increasing distance from the host star and from the Solar System. For Jupiter twins, the noise counts from zodiacal light are within a factor of 2 of those from the planet itself.}

% In Figure~\ref{fig:stddev_astro} we show the noise contributions of different simulated sources in the planet aperture, normalized to the standard deviation of the planet counts. Because the relative contributions of different noise sources are all within roughly one order of magnitude, we plot their contributions on a linear scale and only show the 84th percentile of each distribution as a function of heliocentric distance. In the case of Jupiters in the HZ, the planet's own photon noise is dominant over other sources at every distance; at distances $\lesssim$~2.5 pc and $\gtrsim$~12 pc, speckle residuals in the 84th percentile become comparable to the photon noise contributed by a Jupiter twin, but in the median case the planet's photon noise dominates here as well.

\begin{figure}
\begin{center}
\includegraphics[width=\linewidth]{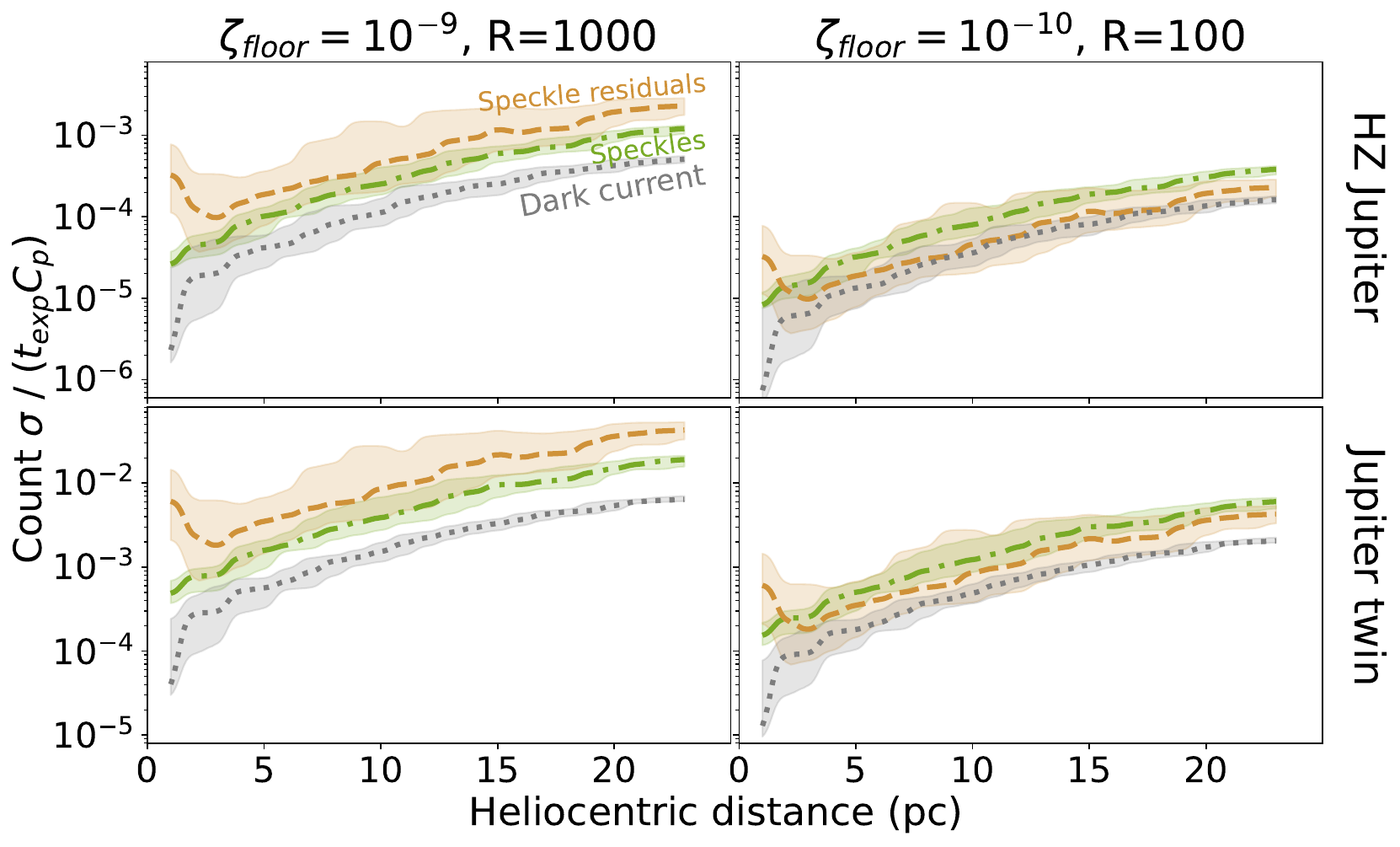}
\end{center}
\caption 
{ \label{fig:stddev_inst}
\rev{Similar to Figure~\ref{fig:stddev_astro}, but for noise sources that are functions of instrument properties and postprocessing. Counts for speckles and dark current are photon noise, but speckle residual counts are given by the total speckle counts times the gain $g$, in this case set to $10^{-2}$. High speckle intensity and high dark current are grouped together to reduce overlap between their respective counts; in our simulations, these sources are varied independently. Data are binned in 1 pc increments with a bin width of 4 pc, then smoothed with a Gaussian filter. Colored lines trace the median values of each bin, and shaded regions are bounded by the 16th and 84th percentiles.}
} 
\end{figure}

\rev{In contrast, Figure~\ref{fig:stddev_inst} shows the contributions of noise sources that depend more sensitively on the observing instrument, i.e. the coronagraph and detectors. Also shown here is the noise contribution from residual speckles. It is important to note that these noise counts are not Poisson distributed, but scale directly with the speckle intensity at the planet location. This means they have a different scaling with both integration time as well as starlight suppression; this leads correlated residuals to exceed the fundamental Poisson noise of the speckles at $\zeta_{floor}=10^{-9}$.}

\rev{Figure~\ref{fig:stddev_inst} also shows the noise contribution of dark current at spectral resolutions of R~=~100 and 1000. The number of dark current counts scales directly with the number of detector pixels, and by extension the spectral resolution (for a fixed bandpass); a tenfold increase in the spectral resolution therefore leads to an increase in the dark current noise contribution by a factor of roughly 3. In our simulated 10-hour observations, the dark current Poisson noise at R~=~1000 doesn't exceed 0.8\% of the total planet counts for HZ Jupiters or Jupiter twins, but at higher spectral resolutions (R~$\gtrsim$~10,000) detector noise may become comparable to speckle noise.}

\begin{figure}
\begin{center}
\includegraphics[width=\linewidth]{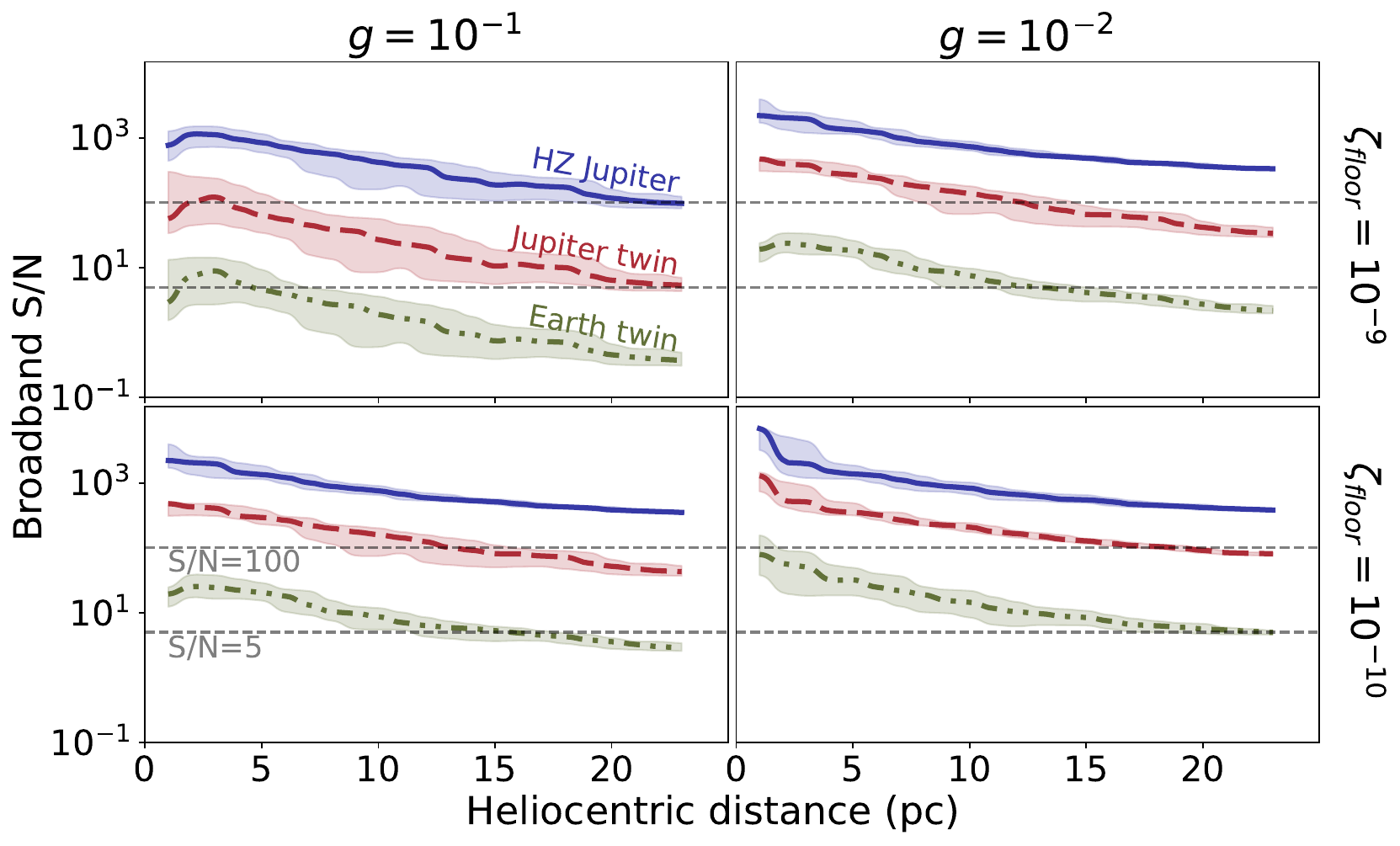}
\end{center}
\caption 
{ \label{fig:snr_v_dist_exp}
Detection S/N of gas giants in simulated HWO observations as a function of heliocentric distance. \rev{Planets are observed in a 20\% bandpass centered at 550nm, and an integration time of 10 hours is used.}
% Plots along the top row show broadband template matching S/N (Eq.~\ref{eq:snr_tm}) on their y axes; plots along the bottom row show S/N per bin (Eq.~\ref{eq:snr_bin}). The left column shows S/N of observations with 10 hour integrations, and the right column shows the same for 400 hours of integration.
\rev{Data are binned in 1 pc increments with a bin width of 4 pc, then smoothed with a Gaussian filter. Colored lines trace the median values of each bin, and shaded regions are bounded by the 16th and 84th percentiles.
}  
} 
\end{figure}

In Figure~\ref{fig:snr_v_dist_exp} we show the broadband S/N (see Section~\ref{sec:snr_bb}) of HZ Jupiters\rev{,} Jupiter twins\rev{, and Earth twins} around ExEP target stars. We find that \rev{HWO can only detect Earth-like planets around the furthest ExEP stars in $\lesssim$10 hours if the coronagraph can meet or exceed a raw starlight suppression factor of $10^{-10}$, and the post-processing gain (i.e. PSF stability) reaches or exceeds $10^{-2}$. This implies a final residual-speckle-to-star flux ratio of $10^{-12}$. By contrast, Jupiter twins are detectable in 10 hours around almost every ExEP star, even in the pessimistic scenario where HWO struggles to detect Earth twins beyond 5 pc. Improving raw starlight suppression seems to yield a similar improvement to the S/N distributions of all three planet types as improving post-processing gain, implying ongoing developments in both of these areas could significantly impact the yield of the mission.}

\rev{In the scenario where exo-Earths are detectable around all ExEP stars in 10 hours,} Jupiter-like exoplanets around all target stars in our simulation have broadband detection S/N~$\gtrsim$~80 for exposure times of 10 hours, even at 5 EEID. For HZ Jupiters these S/N values are a higher by a factor of between 3 and 5, with detections nearer than $\sim$~8 parsecs exceeding S/N~$=$~1000. At the contrasts expected of HWO’s coronagraphic instrument, HZ Jupiters are brighter than all other sources of noise by more than an order of magnitude. Therefore, in all the systems where HWO spends at least one visit searching for exo-Earths, HWO should be able to find gas giants at quadrature phase between 1 and 5 EEID in the same planetary system.

% \begin{figure}
% \begin{center}
% \includegraphics[width=\linewidth]{figs/gas_giant_snr_v_dist_smooth_fourpanel.pdf}
% \end{center}
% \caption 
% { \label{fig:snr_v_dist_exp}
% Detection S/N of gas giants in simulated HWO observations as a function of heliocentric distance. Plots along the top row show broadband template matching S/N (Eq.~\ref{eq:snr_tm}) on their y axes; plots along the bottom row show S/N per bin (Eq.~\ref{eq:snr_bin}). The left column shows S/N of observations with 10 hour integrations, and the right column shows the same for 400 hours of integration. \rev{Data are binned in 2 pc increments with a bin width of 5 pc. Colored lines trace the median values of each bin, and shaded regions are bounded by the 16th and 84th percentiles.}  
% } 
% \end{figure}

After 400 hours of integration, Jupiter twins should have broadband S/N in excess of 300, and at heliocentric distances $\lesssim$13 parsecs should have S/N $\gtrsim$1000. If a Jupiter-like planet is detected in the same system as an exo-Earth that is selected for follow-up characterization, a gas giant that falls within the dark hole may have spectral observations with the same exposure time (though an asymmetric dark hole \cite{kasdin_shaped_2007,por_optimal_2017} and/or a slit/fiber-fed spectrograph may make this more difficult). We discuss spectral characterization of gas giants more in Section~\ref{sec:atmo}. Because gas giants orbiting in the HZs of their host stars would be unlikely have nearby terrestrial planets in stable orbits \cite{painter_astrometric_2025}, such HZ Jupiter planets may not be prioritized for long-duration spectroscopic follow-up with HWO. However, in Section~\ref{sec:exomoons}, we argue that long-duration monitoring of HZ Jupiters could enable the detection of habitable exomoons and further enable HWO's characterization mission.

\section{Atmospheric Molecular Detection} \label{sec:atmo}

In this section, we present \rev{spectral characterization S/N values for the gas giant planets in our simulations. Specifically, we show the average S/N per resolving element of each simulated planet's spectrum ($\langle$S/N$_i\rangle$, see Eq.~\ref{eq:snr_bin}) as well as the template matching S/N (S/N$_{mol}$, see Eq.~\ref{eq:snr_mol}) of molecular absorption features in their spectra, as a function of different observatory parameters.}

\rev{In Section \ref{sec:detection}, we simulate broadband detection in visible light to leverage (1) high planet albedo and (2) a more favorable IWA. For molecular detection, the availability of deep absorption features in the planet spectrum must also factor into the choice of spectral bandpass. We therefore simulate HWO spectroscopic observations of gas giants at a central wavelength $\lambda_c$~=~1000~nm with a 20\% bandwidth; both the HZ Jupiter and Jupiter twin model albedo spectra exhibit deep absorption features from \methane{}, \water{}, and \ammonia{} in this wavelength range (see Figure~\ref{fig:mol_templates}), making it an ideal bandpass to simulate and potentially well suited to observation.}

% Studying the atmospheres of giant planets is a powerful tool for understanding how planets form from the protoplanetary disks orbiting young stars.
% Numerous studies have compared the carbon-to-oxygen ratio (C/O) of hot Jupiters \cite{madhusudhan_high_2011,line_systematic_2014} and directly imaged substellar companions \cite{lavie_helios-retrieval_2017,molliere_retrieving_2020,xuan_are_2024} to their host stars/the Sun; the goal of these studies has been to investigate the degree to which the compositions of these gas giants are fractionated.
% C/O ratios deviating from stellar/solar would imply preferential accretion of metal-enriched solids in the disk, followed by runaway accretion of disk material with near-stellar composition \cite{oberg_effects_2011}.
 
% Furthermore, atmospheric enhancement of nitrogen in the form of \ammonia{} could imply that a planet accreted nitrogen in solid form beyond the N$_2$ ice line, as has been suggested for Jupiter \cite{bosman_jupiter_2019, oberg_jupiters_2019}. \rev{\textcolor{red}{More text here}}

\subsection{S/N per resolving element} \label{sec:atmo_bin}

\begin{figure}
\begin{center}
\includegraphics[width=\linewidth]{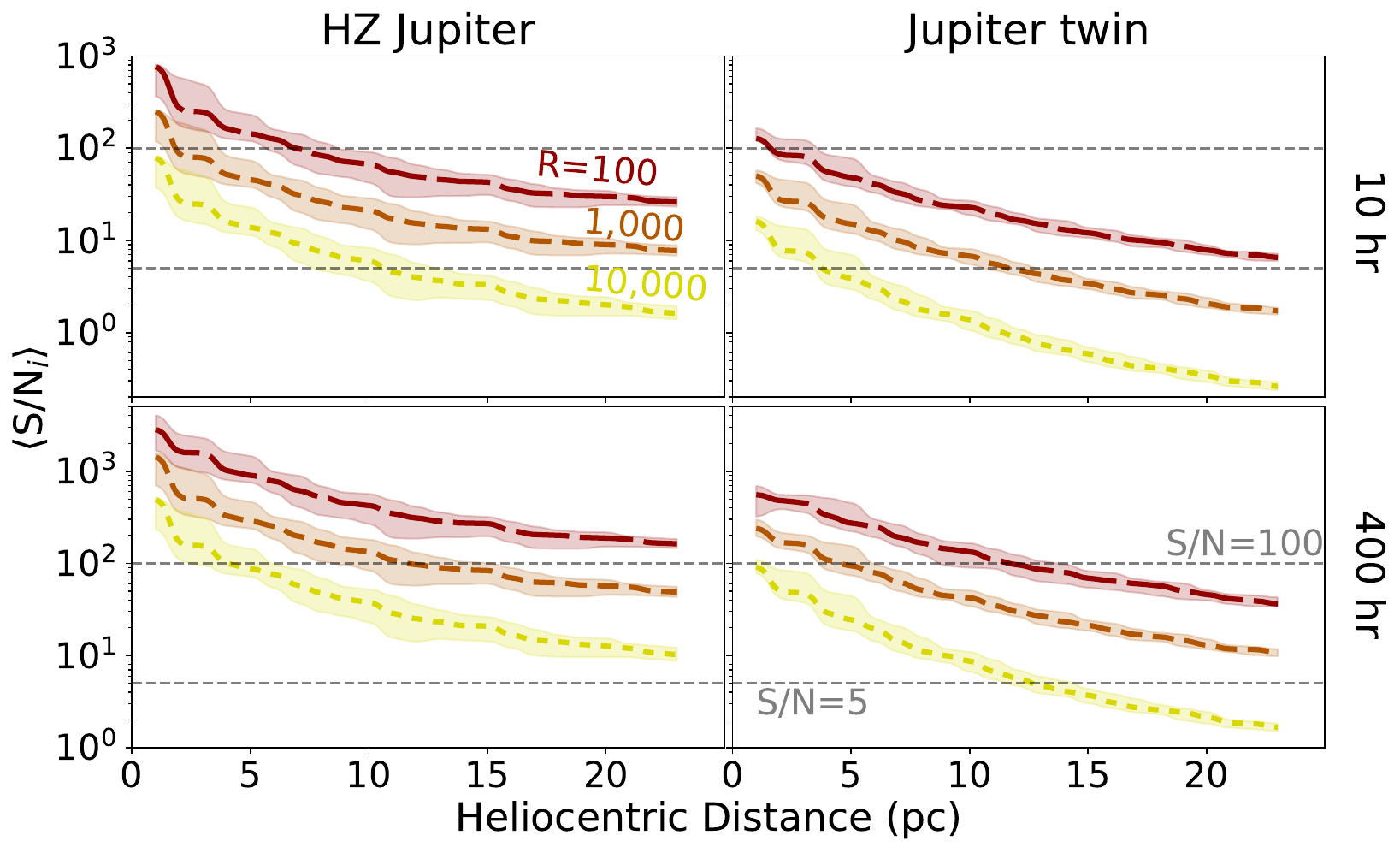}
\end{center}
\caption 
{ \label{fig:bin_snr_res_exp}
 \rev{Mean S/N per spectral resolving element (Eq~\ref{eq:snr_bin}) for each type of simulated planet as a function of spectral resolution and integration time. Panels in the left column correspond to HZ Jupiters, and those in the right column correspond to Jupiter twins. Integration times for the top and bottom rows are 10 and 400 hours, respectively. Each line style/color corresponds to a different spectral resolution. Data are binned in 1 pc increments with a bin width of 4 pc, then smoothed with a Gaussian filter. Colored lines trace the median values of each bin, and shaded regions are bounded by the 16th and 84th percentiles.}
} 
\end{figure}

\rev{In Figure~\ref{fig:bin_snr_res_exp}, we show the dependence of $\langle$S/N$_i \rangle$ on integration time and spectral resolution for HZ Jupiters and Jupiter twins. Studies simulating atmospheric retrievals on exo-Earth \cite{damiano_reflected_2022-1,feng_characterizing_2018, damiano_effects_2025} and gas giant \cite{lupu_developing_2016,nayak_atmospheric_2017,hu_information_2019,damiano_multi-orbital-phase_2020} reflected light spectra use $\langle$S/N$_i\rangle$ to parameterize the amount of noise in their synthetic data. We present $\langle$S/N$_i\rangle$ values for simulated gas giants to inform the results of such studies, and to place constraints on the achievable precision of atmospheric abundance estimates with HWO.}

\rev{For a fixed central wavelength and bandpass, increasing the spectral resolution divides the light of the planet and other astrophysical sources among a larger number of detector pixels, increasing the relative contribution of shot noise to the counts in each resolving element. For nearby gas giants, this effect alone is sufficient to explain the decrease in $\langle$S/N$_i \rangle$ with increasing spectral resolution, where a factor of 10 increase in spectral resolution leads to a factor of $\sim10^{1/2}$ decrease in $\langle$S/N$_i \rangle$. Adding detector pixels also increases the instrument's dark current count rate, but at R~=~100 dark current Poisson noise is negligible for all simulated gas giants (see Figure~\ref{fig:stddev_inst}). For distant Jupiter twins, the dark current from a high-resolution detector has a greater impact; increasing R from 1000 to 10,000 leads to a 6-fold decrease in $\langle$S/N$_i \rangle$.}

\rev{In a single 10-hour spectroscopic observation, HWO can take R~=~1000 near-infrared spectra of HZ Jupiters with $\langle$S/N$_i \rangle\geq$~5 around even the most distant ExEP targets. For HZ Jupiters nearer than 10~--~12 pc, HWO should be able to achieve $\langle$S/N$_i \rangle\geq$~5 at R~=~10,000. Similar results are possible for Jupiter twins at integration times of 400 hours.}

\rev{Spectral resolutions of order $\sim10^3$ are needed to probe the abundances of compounds like sodium and potassium, which have narrow absorption features. These elements are important for constraining the mass of solids accreted by forming gas giant planets. Increased spectral resolution also allows the spectral lines of multiple different compounds to be resolved and disentangled from each other, improving the overall accuracy of abundance estimates from atmospheric retrievals. For exo-Earths, moderate-to-high-resolution spectroscopy may be necessary to detect the presence of biosignatures \cite{ruffio_characterizing_2026}.} 

% \rev{At R~=~1000 and 10 hours of integration per planet, $\langle$S/N$_i \rangle\geq$~5 for all simulated HZ Jupiters, with some nearer targets exceeding $\langle$S/N$_i \rangle$~=~100.}

\subsection{Template matching S/N} \label{sec:atmo_mol}

\begin{figure}
\begin{center}
\includegraphics[width=\linewidth]{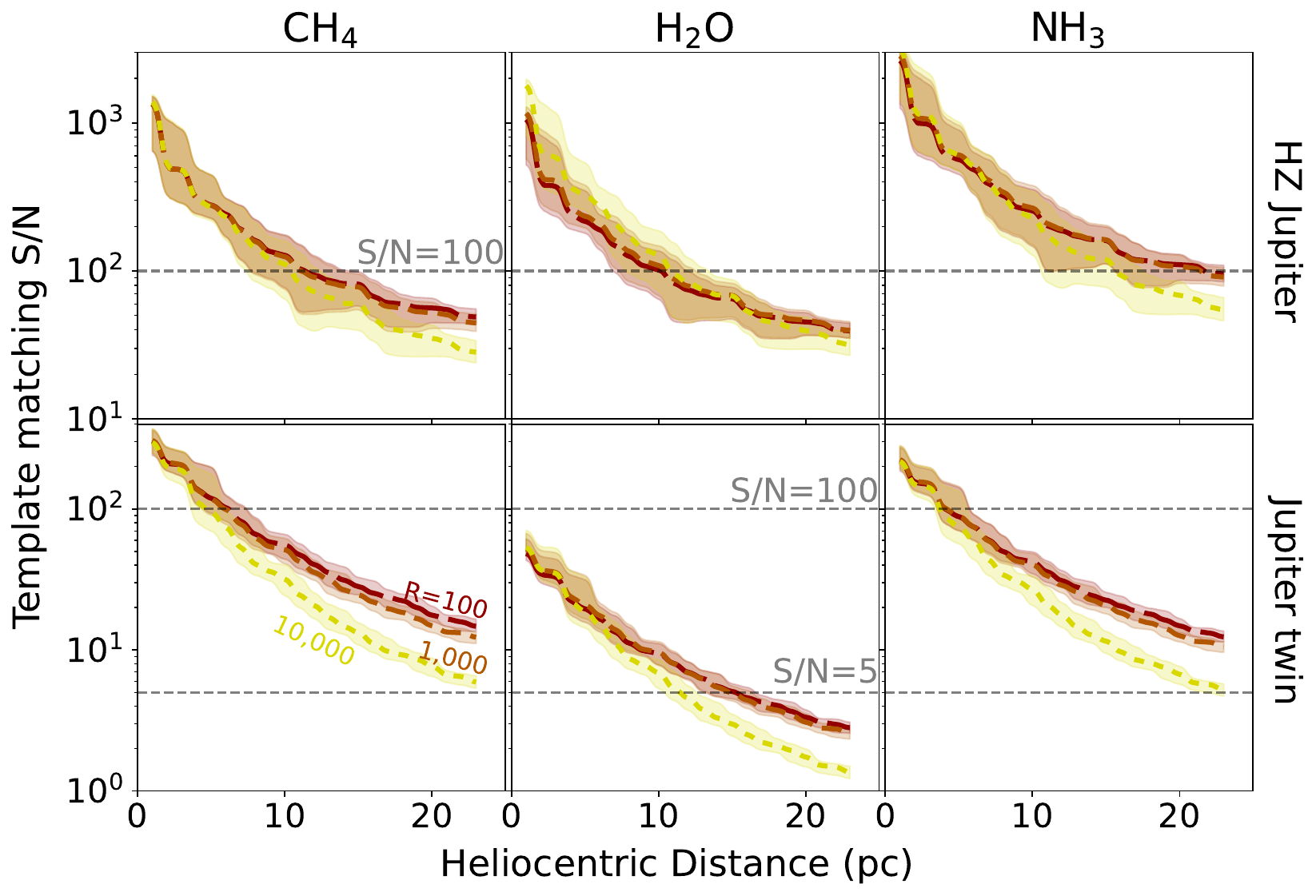}
\end{center}
\caption 
{ \label{fig:mol_snr}
\rev{Template matching S/N of molecular absorption features in the spectra of gas giants in our simulations after 10 hours of integration. The top row shows molecular detection S/N for HZ Jupiters while the bottom row displays that of Jupiter twins. From left to right, each column corresponds to methane (\methane{}), water (\water{}), and ammonia (\ammonia{}). Different line styles/colors show different spectral resolutions. Data are binned in 1 pc increments with a bin width of 4 pc, then smoothed with a Gaussian filter. Colored lines trace the median values of each bin, and shaded regions are bounded by the 16th and 84th percentiles.}
} 
\end{figure}

\rev{Figure~\ref{fig:mol_snr} shows template matching S/N values of \methane{}, \water{}, and \ammonia{} absorption in simulated gas giant atmospheres. Individual molecular templates and the S/N of their detection are defined in greater detail in Section~\ref{sec:snr_mol}. Template matching S/N has the advantage of incorporating the entire spectrum and its uncertainties in its calculation, similarly to the likelihood estimation used by atmospheric retrievals when fitting atmospheric parameters like molecular abundances. This methodology also allows systematics like speckle residuals to be incorporated into the noise model by way of a covariance matrix; this is crucial, as correlations in the noise counts of nearby pixels can masquerade as genuine spectral features of similar width, potentially degrading a spectrum's information content. To maximize the scientific constraining power of future retrievals using HWO data, it is therefore imperative that we assess the spectral resolution trade space by modeling the planetary signal and sources of noise as accurately as possible.}
% At single-visit integration times of 10 hours, we find that \water{} and \methane{} are detectable in the atmospheres of all gas giants in our simulation, and that \ammonia{} is detectable in the atmospheres of Jupiter twins out to $\sim$13 parsecs using template matching at a spectral resolution of $>$~100.

% \begin{figure}
% \begin{center}
% \includegraphics[width=\linewidth]{figs/mol_snr_v_dist_smooth_instellations.pdf}
% \end{center}
% \caption 
% { \label{fig:mol_snr_instellation}
% Spectroscopic S/N for Jupiter-like planets at 1 (left) and 5 (right) EEID, assuming a spectral resolution of 100 and an integration time of 10 hours. For the three molecules (H2O, CH4, NH3), values are template matching S/N shown in Equation~\ref{eq:snr_mol}, assuming a correlation length scale of 10 nm. \rev{Data are binned in 2 pc increments with a bin width of 5 pc. Colored lines trace the median values of each bin, and shaded regions are bounded by the 16th and 84th percentiles.}
% } 
% \end{figure}

% Interestingly, the S/N of ammonia in gas giant atmospheres, and of all three molecules in the atmospheres of HZ Jupiters, is less than the average S/N per bin (shown as the dotted orange line). This is due to the effects of correlated noise, which we include in our simulations (see Section~\ref{sec:snr_mol}). At low spectral resolution, correlated noise can mimic absorption from atmospheric molecules. However, by increasing the instrument's spectral resolution, we can make unresolved absorption features deeper and narrower without changing the length scale of correlated noise. We therefore explore how spectral resolution influences S/N in Figures~\ref{fig:vol_mol_snr_res} and~\ref{fig:nh3_snr_res}.

\rev{The template matching S/N shown in Figure~\ref{fig:mol_snr} has a very different dependence on spectral resolution than the S/N per resolving element in Figure~\ref{fig:bin_snr_res_exp}. For all molecules in HZ Jupiter atmospheres, S/N is essentially unchanged between R~=~100 and R~=~1000; this is the expected behavior when the planet's own uncorrelated photon noise dominates the noise budget. However, there is reason to think that correlated speckle residuals (flux ratio of $10^{-12}$ with the host stars) is having an effect: at R~=~10,000, S/N$_{H_2O}$ is actually 50\% higher for the nearest HZ Jupiters. This demonstrates how spectroscopy at moderate-to-high R can be used to more reliably classify spectral features as noise or true indicators of atmospheric composition and physics.}

\rev{For Jupiter twins, and HZ Jupiters further from the Solar System, template matching S/N decreases more noticeably at R~=~10,000. The low count rate from these more distant targets and the large number of pixels in a high-resolution detector significantly increases dark current’s relative contribution to the noise budget. However, the change between R~=~100 and R~=~1000 is still relatively small, and in almost every case the template matching S/N remains above 5 at all spectral resolutions. \water{} in the atmospheres of Jupiter twins stands as an exception; in our spectral models, the water vapor on these planets has almost entirely condensed, leaving the atmosphere above the \ammonia{} cloud deck almost entirely devoid of water. This leads to a lack of significant \water{} absorption past $\sim$~950nm in the Jupiter twin model spectrum (see Figure~\ref{fig:mol_templates}).}

To reliably distinguish overlapping absorption features of \water{}, \methane{}, \ammonia{}, and other molecules would require spectral resolution of a few $10^3$. Furthermore, for detecting narrow atomic lines like those from sodium and potassium, moderate-to-high spectral resolution is a necessity. These and other refractory species are key tracers of solid accretion onto forming gas giants because they occur in solid phase throughout the vast majority of the protoplanetary disk \cite{schneider_how_2021,turrini_tracing_2021}. [\citenum{cahoy_exoplanet_2010}] find that there should be significant optical absorption features from these refractories in the light reflected from gas giants at a wide variety of instellations, which HWO should be able to detect at sufficiently high spectral resolution. \rev{However, our models do not include these additional species, and we are not aware of any high-resolution reflected-light spectral models that do; we therefore leave this analysis to future work.}

[\citenum{ruffio_jupiter-like_2026}] use the James Webb Space Telescope's NIRSpec instrument to collect R~$\sim$2,700 spectra of HR8799 c, d, and e at 3---5~$\mu$m. With high-pass filtered template matching S/N values ranging from 79 to 134, the authors were able to constrain the abundances of carbon, oxygen, and sulfur, as well as the $^{12}$CO/$^{13}$CO and C$^{16}$O/C$^{18}$O isotopologue ratios. [\citenum{xuan_are_2024}] use Keck KPIC spectra (R~$\sim$35,000) from $\sim$~2---2.5~$\mu$m to estimate the $^{12}$C/$^{13}$C isotopic ratios of 8 directly imaged substellar companions (S/N between 7 and 80). Spectral resolutions of~$\sim$~10,000 or greater could resolve spin-broadened lines down to $v\sin i \approx 30$km/s, providing an estimate of a gas giant's line-of-sight rotation speed and allowing atmospheric retrievals to more accurately estimate the abundances of absorbing molecules. \rev{Yield} and retrieval simulations using high-resolution model spectra will be needed \rev{in the future} to explore the optimal spectral resolution for characterizing large \rev{numbers} of gas giants and habitable planets \cite{ruffio_characterizing_2026}.

%\textbf{JB: probably need to add more explanation to explain that we might also need high R to disentangle NH3 from H2O/CH4 features for example. Also from Figure 1, looks like NH3 is a lot shallower in Jupiter than the models. Also could we comment on the possibility to detect K and Na TBD? and Maybe isotopologues? even if we don't have the models? possibly in the discussion section TBD}
% Even if only one gas giant exists around one of these targets, and that the S/N of this detection increases with increasing spectral resolution up to the resolution limit of our models. Increasing the integration time from 10 hours to 100 hours more than triples this S/N for the nearest systems, but the farthest are not in the photon-noise-limited regime, so the effect is less pronounced, especially at high spectral resolution. We note that, for other molecules (for which our models are higher resolution) the optimal spectral resolution (assuming detector noise from EMCCD technology that may be achievable by the time construction on HWO begins), the spectral resolution that maximizes the S/N is FGH.

% \begin{figure}
% \begin{center}
% \includegraphics[width=0.8\linewidth]{figs/snr_v_dist_exptime10_lambda650_override_local_starlight_flux_ratio1e-11_ppFact_Char0.01_eecoef5.0_renyuJupiter_5AU_beta0.5.pdf}
% \end{center}
% \caption 
% { \label{fig:nh3_5au}
% NH3 detection SNR at different spectral resolutions, 10hr integration, 5EEID, 650nm TODO: 5EEID
% } 
% \end{figure}

\section{Photometric Variability Detection} \label{sec:lightcurve}
 Similar to Section~\ref{sec:detection}, in this section we present broadband S/N ratios for simulated Jupiter-like planets, but with a focus on photometric variability studies. We therefore explore shorter integration times of 5 minutes and 1 hour.
 
\begin{figure}
\begin{center}
\includegraphics[width=\linewidth]{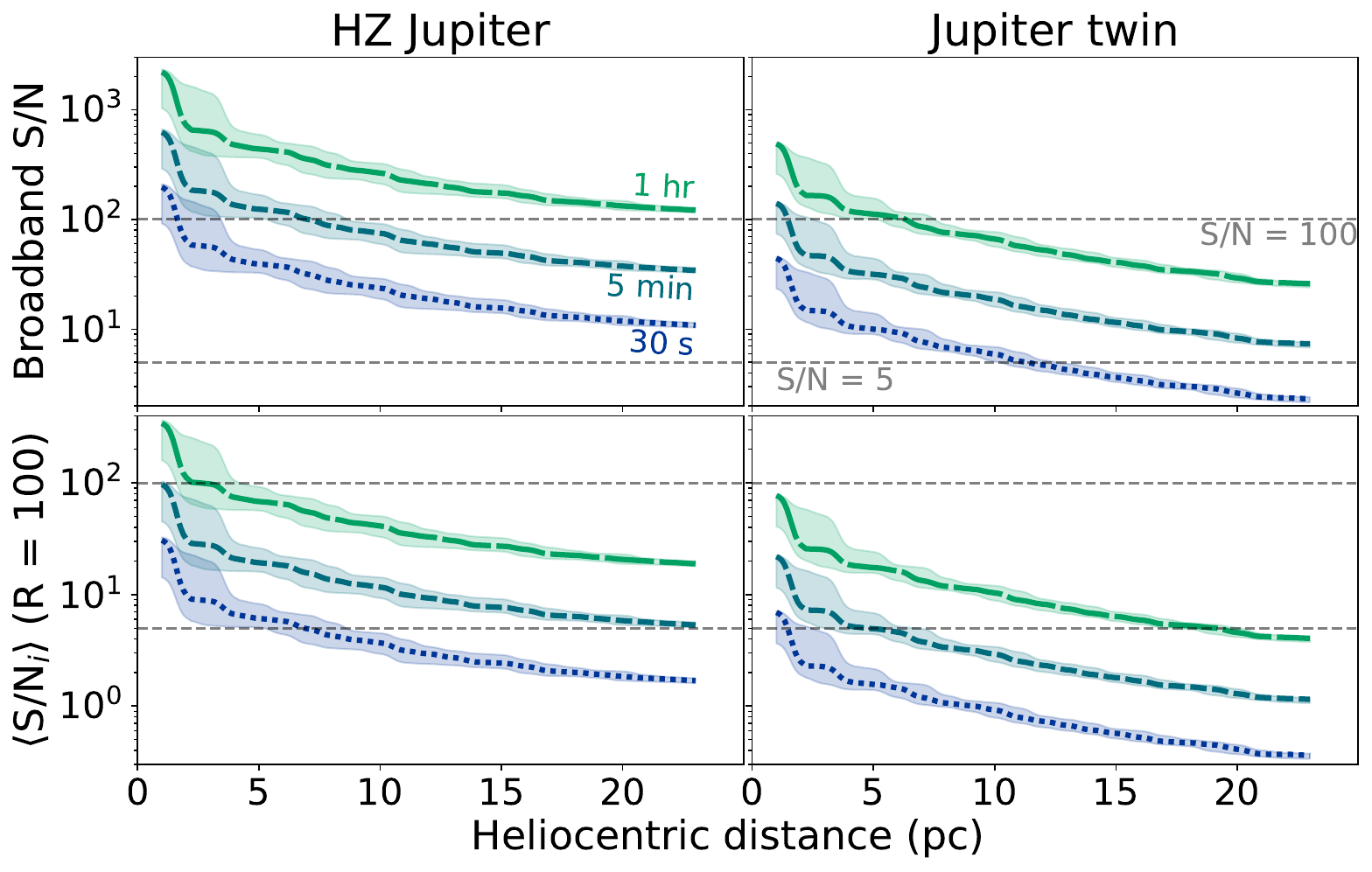}
\end{center}
\caption 
{ \label{fig:cadence}
Similar to Figure~\ref{fig:snr_v_dist_exp} but for integration times of \rev{30 seconds,} 5 minutes\rev{,} and one hour. A horizontal dashed line has been added at S/N=100 to show where gas giant photometric precision equals 1\%, enabling studies of rotational variability comparable to Jupiter's \cite{ge_rotational_2019}. \rev{Data are binned in 1 pc increments with a bin width of 4 pc, then smoothed with a Gaussian filter. Colored lines trace the median values of each bin, and shaded regions are bounded by the 16th and 84th percentiles.}
} 
\end{figure}

 Figure~\ref{fig:cadence} shows the broadband detection S/N and the S/N per resolving element of simulated gas giant planets in short exposures. \rev{The noise sources included in these simulations are those listed in Section~\ref{sec:spectrum} and Equation~\ref{eq:specvec}.} We find that Jupiter-like exoplanets, including Jupiter twins, around all target stars in our simulation have broadband detection S/N~$>$~5 after a single 5-minute exposure. The S/N values of HZ Jupiters exceed those of Jupiter twins by factors similar to those found in Section~\ref{sec:detection}. To ensure sufficient sampling of exomoon light curves, we therefore choose to set the exposure time of each observation in Section~\ref{sec:detection_photo} to 5 minutes. However, the use of template-matching S/N for photometric event detection (see Section~\ref{sec:photvar}) means that the results should apply for a wide range of individual exposure times.

 After 1 hour of integration, all HZ Jupiters, as well as Jupiter twins closer than $\sim$~8 pc to the Solar System, are detectable with S/N~$\gtrsim$~100, implying a 1$\sigma$ photometric senisitivity of 1\%. [\citenum{ge_rotational_2019}] find that, over the course of Jupiter's $\sim$~10 hour rotation period, its flux in reflected visible light varies with an amplitude of roughly 1\%. Similarly, [\citenum{akhmetshyn_mapping_2025}] use JWST NIRISS data to show roughly 1\% flux variability of the thermal emission of a low-mass, isolated brown dwarf. Achieving better than 1\% sensitivity in the broadband flux of a gas giant in $\sim$1 hour allows for measurement of planetary rotation rates, studies of temporal atmospheric variability, and potentially the detection of moons or rings orbiting these planets (the latter is explored in Section~\ref{sec:exomoons}).

 For HZ Jupiters, the S/N per resolving element at a spectral resolution of 100 is generally greater than 5 after a five minute exposure, and in 1 hour $\langle$S/N$_i \rangle$ can exceed 100 for nearby HZ Jupiters. This could potentially allow for wavelength-dependent variability studies revealing time-resolved compositional and structural changes in exoplanet atmospheres, similar to past studies of Solar System gas giants \cite{ge_rotational_2019, blake_saturns_2023}.

\section{Detection and Characterization of Exomoons} \label{sec:exomoons}

\subsection{Exomoon Photometric Detection} \label{sec:detection_photo}

As shown in Section~\ref{sec:detection}, Jovian planets are detectable in individual 5-minute exposures with HWO. Stacking these exposures together can increase the S/N of the planet signal as well as molecular absorption features in its spectra, but keeping them separate allows us to construct a photometric time series of the planet, where the error on each photometry point is given by $f_p / \mbox{S/N}$. Such time series can be phase-folded to reveal periodic patterns, and can be fit with models of planet-moon eclipse and obscuration events to reveal the existence of satellites.

We describe in Section~\ref{sec:methods} how we simulate the photometric effects of exomoons on planet light curves. In this section, we present the results of our simulations. \rev{We take as the fiducial case an Earth-radius moon orbiting a Jupiter-radius, Jupiter-mass planet with the same orbital period and semi-major axis as Io around Jupiter. If an entire primary eclipse event in such a moon-planet system were captured in an HWO light curve with a 5 minute cadence, the photometric event strength $H=0.073$, requiring that the planet have broadband detection S/N~$\geq$~68 in order for a single such event to be detected at S/N~=~5. Because only the light of the planet itself is being blocked, the albedos of the planet and moon do not matter as long as $R_{\leftmoon}/R_p \ll A_{g,p}/A_{g,\leftmoon}$, where $R_{\leftmoon}$ and $A_{g,\leftmoon}$ are the moon's radius and albedo, and $R_p$ and $A_{g,p}$ are those of the planet. It can be shown using Equation~\ref{eq:photo_strength} that an event's strength $H$ is directly proportional to the overall strength of the signal, which in this case scales with the square of the ratio of the moon and planet’s radii. Further, the strength of multiple events observed consecutively can be calculated by taking the quadrature sum of each event's individual strength (this does not work if the events overlap with each other in time.) We can therefore approximate a more general formula for the photometric strength of a primary eclipse:}
\begin{equation} \label{eq:prim_ec_strength}
    H_{pe} \approx 0.073\left( \frac{R_{\leftmoon}/R_p}{R_{\oplus}/R_J}\right )^2 n_{orb}^{1/2}
\end{equation}
\rev{where $H_{pe}$ is the strength of a series of primary eclipses and $n_{orb}$ is the number of consecutive events seen, i.e. the number of orbits that fall within the observing window.}

\rev{Similarly, the strength of a secondary eclipse in our fiducial example is 0.041. To detect a single secondary eclipse at S/N~=~5, the host planet would need to have a broadband S/N per exposure of~$\geq$~122. This assumes an equal albedo for the planet and moon; unlike in a  primary eclipse, the flux change that occurs during a secondary eclipse is caused by the (lack of) light reflected by the exomoon, requiring the moon's albedo to be specified. We can create a similar expression to Equation~\ref{eq:prim_ec_strength} for secondary eclipses:}
\begin{equation} \label{eq:sec_ec_strength}
    H_{se} \approx 0.041\left( \frac{R_{\leftmoon}/R_p}{R_{\oplus}/R_J}\right )^2 \left( \frac{A_{g,\leftmoon}}{A_{g,p}} \right) n_{orb}^{1/2}.
\end{equation}

\begin{figure}
\begin{center}
\includegraphics[width=\linewidth]{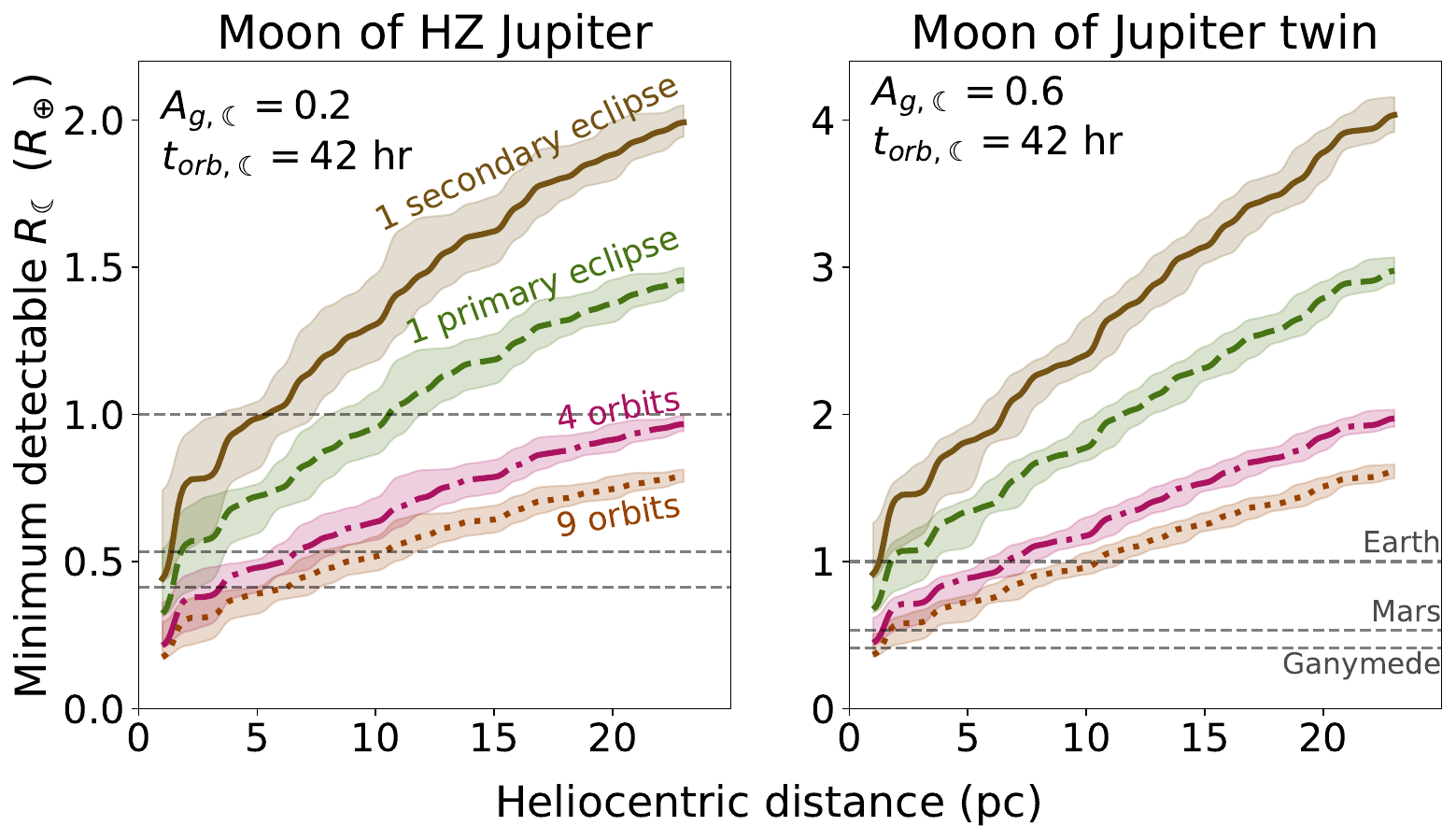}
\end{center}
\caption 
{ \label{fig:minimum_radius}
The minimum radius required for a 5$\sigma$ detection of an exomoon-caused photometric event (see Section~\rev{\ref{sec:exomoon_methods}}). Left: the case of an Earthlike moon orbiting an HZ Jupiter around each target star in our simulation. Right: the case of a moon with an albedo matching Europa's orbiting a Jupiter twin. In both cases the moon has the same orbital period as Io, 42 hours. \rev{Data are binned in 1 pc increments with a bin width of 4 pc, then smoothed with a Gaussian filter. Colored lines trace the median values of each bin, and shaded regions are bounded by the 16th and 84th percentiles.}
} 
\end{figure}

\rev{Adding Equations~\ref{eq:prim_ec_strength} and~\ref{eq:sec_ec_strength} in quadrature gives the strength of a light curve consisting of $n_{orb}$ primary and secondary eclipses. We can further rearrange said expression to solve for the minimum $R_{\leftmoon}$ for which a photometric signal would be detectable at S/N~=~5:}
 \begin{equation} \label{eq:minrad}
     R_{\leftmoon,min} = R_{\oplus} \left( \frac{R_p}{R_J} \right)^{-1} \left( \frac{5}{\text{S/N}_{p,5min}} \right)^{1/2} \left[ n_{orb}(0.073^2 + \left( 0.041\frac{A_{g,\leftmoon}}{A_{g,p}} \right)^2 \right]^{1/4}
 \end{equation}
 \rev{where S/N$_{p,5min}$ is the detection S/N of the host planet in a 5-minute exposure.} Figure~\ref{fig:minimum_radius} \rev{shows $R_{\leftmoon,min}$ for the planets in our simulations}, assuming the annotated event(s) occurred within the observing window. In the case of a moon with an Earth-like albedo around an HZ Jupiter, moons Earth's size or smaller are detectable after a single shadow event out to roughly \rev{10} parsecs.

Equation \ref{eq:minrad} shows that, by increasing the temporal baseline of photometric monitoring, one can probe gas giant exomoons with smaller radii. For a moon with a similar orbital period to Jupiter's moon Io (42 hours), a 400-hour characterization stare with HWO could coincide with 9 orbits, therefore allowing up to 9 shadows and 9 eclipses to be detected. Assuming all of these events are observed and accurately modeled, this can then triple the SNR and significantly reduce the minimum exomoon radius required for detection. In the case where 9 orbits are captured, Earth-sized exomoons are visible around HZ Jupiters orbiting all the stars in the ExEP catalog. Mars-sized moons become detectable at distances less than $\sim$10 parsecs, and moons the size of Jupiter's moon Ganymede can be detected at distances closer than $\sim$7 parsecs.

Due to decreased instellation, exomoons around Jupiter twins are more difficult to detect. This means that Earth-sized exomoons could only be detected around them from a single shadow if their host star was one of the nearest to the Sun. On the other hand, these objects are likely to have much icier surfaces than an Earth-like moon\rev{; secondary eclipses of exomoons orbiting Jupiter twins may therefore contribute significantly more to a light curve than for HZ Jupiters. The minimum radii at which moons orbiting simulated Jupiter twins would be detectable are shown in the right panel of Figure~\ref{fig:minimum_radius}. We find that, while the primary eclipse still provides the tightest radius constraints, the secondary eclipse reaches comparable S/N, and with 9 orbits' worth of photometric monitoring it may still be possible to detect Earth-sized, icy moons around Jupiter twins up to $\sim$~10 pc away.}

% We show that single transit events are detectable out to a distance of S parsecs for a moon comparable in size to Ganymede ($0.4R_\oplus$) orbiting a Jupiter twin (see Figure~\ref{fig:shadow_SNR}). Increasing the radius of the icy moon to $1R_\oplus$ increases the maximum detectable distance to T parsecs.

% \begin{figure}
% \begin{center}
% \includegraphics[width=\linewidth]{figs/ganymede_photom_snr_v_dist_lessoptimistic.pdf}
% \end{center}
% \caption 
% { \label{fig:eclipse_SNR}
% The detection signal-to-noise ratio of photometric events as a function of heliocentric distance. The planet is assumed to have a radius of 1$R_J$ and to be in quadrature phase.\textcolor{red}{TODO: fix wavelength label (1385nm)}
% } 
% \end{figure}

% \begin{figure}
% \begin{center}
% \includegraphics[width=\linewidth]{figs/moon_photom_snr_v_dist_lessoptimistic.pdf}
% \end{center}
% \caption 
% { \label{fig:shadow_SNR}
% The detection signal-to-noise ratio of photometric events as a function of heliocentric distance. The planet is assumed to have a radius of 1$R_J$ and to be in quadrature phase.
% } 
% \end{figure}

\subsection{Exomoon Spectral Characterization} \label{sec:moonspec}

In principle, the reflected light signal we receive from an Earth-like exomoon is the same as that from an isolated exo-Earth. The parent planet introduces two complicating factors: additional photon noise and spectral confusion from its own spectral features.  While the additional photon noise cannot be mitigated, there are ways to circumvent the blending of planetary and lunar spectral features. One way is to simply argue that a particular spectral feature-- say, that of molecular oxygen-- should not exist in the atmosphere of a gas giant. This could be done, for example, using multi-object retrievals: if a planet plus exomoon model is shown to perform markedly better than a planet-only model when fitting the planet spectrum, this could allow for exomoon detection as well as characterization of its surface and/or atmosphere.

\begin{figure}
\begin{center}
\includegraphics[width=\linewidth]{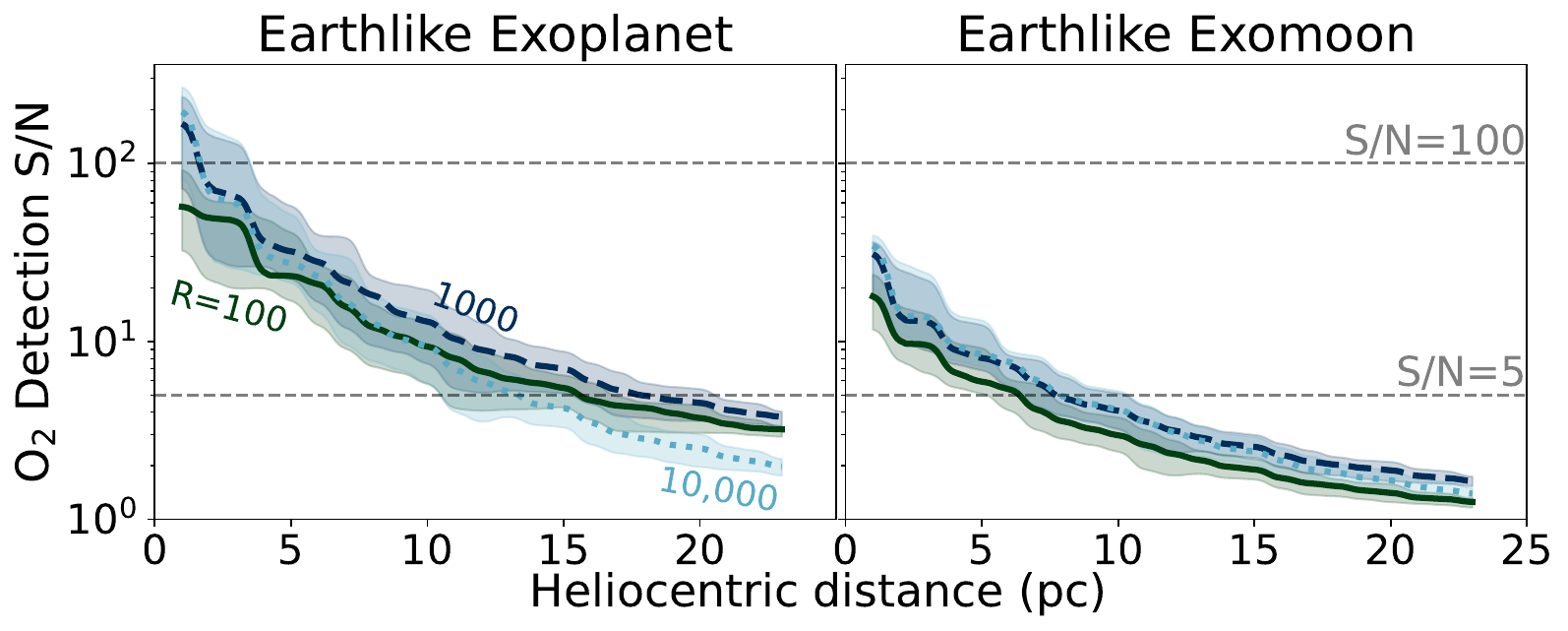}
\end{center}
\caption 
{ \label{fig:snr_earth_moon_O2}
Atmospheric O$_2$ detection SNR in the atmosphere of (left) an Earth-like planet at 1 EEID and (right) and Earth-like moon orbiting an HZ Jupiter planet at 1 EEID. Different shades of each color represent different spectral resolutions, ranging from 100 to \rev{10,000}. The S/N = 5 detection limit is plotted with a dashed horizontal line. \rev{Data are binned in 1 pc increments with a bin width of 4 pc, then smoothed with a Gaussian filter. Colored lines trace the median values of each bin, and shaded regions are bounded by the 16th and 84th percentiles.}
} 
\end{figure}

Another way is to use radial velocity differences between the planet and moon to distinguish their contributions to the spectrum. Gas giant planets will typically have broader absorption lines than terrestrial planets due to their increased surface pressures and $v\sin i$ values, and over the course of a moon's orbit its spectral lines will shift relative to the planet's ($\sim$~17~km/s for an Io-like moon). At R~$\gtrsim$~10,000, these effects could be used to disentangle the molecular absorption features of the moon from those of the planet, allowing us to detect molecules like water even though the atmospheres of many gas giants contain water as well. We show the results of our simulations of exomoon oxygen detection in Figure~\ref{fig:snr_earth_moon_O2}. These results show that it may be easier to detect oxygen on Earthlike moons within 10 pc than for isolated, Earth-like exoplanets at 25 pc, even at a spectral resolution of 10,000. This implies that HWO should also search for habitable worlds orbiting nearby gas giants orbiting in the circumstellar habitable zone.

%\textcolor{red}{WARNING: the following texts is based on extrapolations from new, preliminary simulations, not the following figure. Results should be taken with a grain of salt!} We find that, after 400 hours of integration, oxygen is detectable on exomoons with sizes and atmospheres similar to Earth at distances of $\sim$10 parsecs or less. Figure~\ref{fig:snr_earth_moon_O2} compares this to the S/N of oxygen detection for an isolated exo-Earth using the same methods. Because the contrast floor of the dark hole in our simulations is fixed, while the flux ratio between planet and star varies such that the instellation received by the planet remains constant, isolated Earths around more distant, brighter stars will be more impacted by speckle photon noise. The net effect of this is that the slope of O$_2$ detection S/N as a function of distance is steeper for isolated Earths than it is for Earth-like exomoons, where the photon noise from the host planet essentially always dominates. This means that a hypothetical Earthlike exomoon at 3 parsecs distance might be easier to search for biosignatures than an isolated exo-Earth 10 parsecs or more away.

\section{Discussion} \label{sec:discussion}

\subsection{Gas giant atmosphere characterization}
To inform HWO’s ability to characterize gas giant exoplanets in reflected light, we used EXOSIMS\cite{savransky_wfirst-afta_2015} to simulate HWO observations of Jupiter-sized planets ($1$--$5\,$EEID) around 164 stars\cite{mamajek_nasa_2024} in the solar neighborhood to estimate their achievable S/N.
First, we showed that at quadrature phase, such Jupiter-sized planets are typically detectable in broadband at S/N $>5$ in only five minutes of exposure time (Figure~\ref{fig:snr_v_dist_exp}). 
The planets’ own photon noise dominates over all other sources of noise (exozodi, starlight, etc.) in all but the worst-case scenarios (Figures \ref{fig:stddev_astro} and \ref{fig:stddev_inst}).
Assuming similar detector noise to LUVOIR \cite{the_luvoir_team_luvoir_2019}, the detector noise does not dominate the noise budget at spectral resolution lower than $R<1,000$, which suggests that moderate to high resolution spectroscopy would be particularly powerful for the characterization of gas giant atmospheres with HWO.
In Figure~\ref{fig:mol_snr}, we showed than \ammonia{} could be detectable at high significance at most stellar distances in $>10\,$hours of exposure time. This is similar to the proposed exposure times required to initially detect exoEarths during the first phase of typical survey strategies for HWO  \cite{the_luvoir_team_luvoir_2019,gaudi_habitable_2020}. Therefore, we suggest that using a spectrograph during this initial phase, instead of broadband imaging, could provide detailed atmospheric characterization of most gas giants without additional dedicated observing time. This could significantly increase the science returns of this phase of the mission by enabling HWO's broader exoplanet science goals; namely to better understand the atmospheres of large planets and test planet formation theories.
However, the detailed exploration of giant planet science cases is limited by the availability of atmospheric models, in particular at sufficient spectral resolution ($R>1,000$). This is due, in part, to limitations of the line lists \cite{rothman_hitran2012_2013}. More detailed models of gas giant atmospheres will be needed to explore the transformative potential of moderate to high-resolution spectroscopy for example to detect refractory species and isotopologues, which could provide an important window to understand planet formation. Indeed, recent results with JWST (e.g., \cite{ruffio_jupiter-like_2026}) have shown that $R\sim2,700$ enables detections of elements beyond carbon and oxygen, such as sulfur, nitrogen, and isotopologues that would be otherwise be inaccessible at lower spectral resolution ($R<1,000$). More detailed atmospheric models at higher spectral resolution will also be necessary to explore other science cases including radial velocity measurements \cite{ruffio_radial_2019} and spin-broadening \cite{wang_detection_2021}. Future work should also expand this study to smaller gas giants and super Earths.

\subsection{Occurrence Rates of gas giants} \label{subsec:occurence}
%\textcolor{red}{(Discuss gas giant occurence rate. Will there be enough giant planets given rough exomoon occurence rates? probably comment on known planets right now and future Gaia DR4's impact? Also comment on importance of OWA for 5EEID planets in the solar neighborhood?)}.

\rev{Previous HWO yield studies studies randomly sample from terrestrial planet occurrence rates to estimate the number of planets that HWO can expect to find \cite{kopparapu_exoplanet_2018,morgan_faster_2021,howe_scientific_2024, plavchan_analytic_2024, stark_paths_2024}. In this section, we make a rough estimate of the number of gas giants that HWO could observe and characterize based on known gas giant occurrence rates and the search completeness of previous radial velocity surveys.}

The California Legacy survey \cite{fulton_california_2021} used radial velocity data of 719 stars to estimate the occurrence rates of gas giants orbiting between 0.03 and 30 AU\rev{; they find that, between 1 and 5 EEID, the occurrence rate of gas giants between 0.1 and 20 $M_J$ is roughly 10\%. Assuming that the mass and instellation distributions are independent, about 45\% of those planets are expected to lie in their host stars' HZs (see Figure 11 of [\citenum{fulton_california_2021}]).} We use Figure 7 from [\citenum{fulton_california_2021}] to assign an expected number of gas giants ($0.3M_J\leq M_p \leq 20M_J$, 1 AU~$\leq a_p \leq$~5 AU) to each star in the ExEP target list; we then sum these values for all 164 stars and arrive at a rough expected yield of 23 gas giant exoplanets.
\rev{16} gas giants near or within this range \rev{of instellations} have already been found orbiting 12 ExEP targets, namely: $\epsilon$ Eridani b \cite{thompson_revised_2025}; HD 219134 h \cite{vogt_six_2015, johnson_12-year_2016}; 55 Cancri A d and f \cite{fischer_five_2008};
% HD 147513 b \cite{mayor_coralie_2004};
$\upsilon$ And A c and d \cite{mcarthur_new_2010,piskorz_detection_2017}; 47 Uma b and c \cite{gregory_bayesian_2010}; $\mu$ Ara b, c, and e \cite{gozdziewski_orbital_2022}; Gliese 777 b \cite{giovinazzi_neid_2025}; $\iota$ Hor b \cite{boulkaboul_magnetic_2025,arenou_gaia_2023,zechmeister_planet_2013}; q$^1$ Eri b\cite{bisht_radial_2024}; $\pi$ Men b \cite{huber_20_2022}; and HD 33564 b \cite{galland_extrasolar_2005}. \rev{Of these planets, 3 ($\sim$~19\%) are in their circumstellar HZs ($\upsilon$ And A d, $\mu$ Ara b, and 47 UMa b). This is a significantly lower fraction than that found by [\citenum{fulton_california_2021}], even accounting for the large error bars on their occurrence rate estimates; however, the majority of these 16 planets are in mutli-planet systems, and dynamical stability requirements make it unlikely that multiple gas giants could orbit in a single star's HZ.}
By the time of HWO’s launch, it is likely that precision radial velocity surveys \cite{harada_spores-hwo_2024} and Gaia astrometry will have revealed \rev{many of the remaining} gas giants that have not already been found \cite{brandt_hipparcos-gaia_2021}.

\rev{With 16 gas giants discovered orbiting ExEP target stars, and $\sim$~23 expected to exist, we might naively expect roughly 7 undiscovered gas giants to be orbiting the ExEP target stars. Indeed, this is in agreement with the findings of [\citenum{harada_spores-hwo_2024}], who collected archival radial velocity data on the ExEP target stars to constrain the presence of gas giants that could disrupt the orbits of planets in the stars' HZs. They found that 44 of the ExEP stars have essentially no constraints on orbiting planets; if we assume the 10\% occurrence rate from [\citenum{fulton_california_2021}] applies to these stars, we would expect roughly 4 to 5 undiscovered gas giants to be orbiting around them. Of the remaining 120 stars, roughly 45 have RV completeness values $\lesssim$~50\% for Saturn-mass planets in the HZ, and 17 of those are $\lesssim$~50\% complete to Jupiter-mass planets. [\citenum{fulton_california_2021}] find that, of the gas giants in their survey with masses between 0.1 and 20 $M_J$, roughly 60\% were less massive than Jupiter, and 25\% were less massive than Saturn. Assuming the gas giant mass distribution is independent of instellation, the  occurrence rates for sub-Jupiters and sub-Saturns would be 6\% and 2.5\%, respectively. Applying these occurrence rates to each category of stars, and multiplying them by a conservative conditional detection probability of 50\%, we find that there should be approximately 1 additional undiscovered gas giant among them, bring the total to roughly 5 or 6. This likely represents an underestimate; we assumed that past RV surveys of all of these stars were exactly 50\% complete, but for many of them it will be much less, especially exterior to the habitable zone. We also did not consider planets more massive than Jupiter, which [\citenum{fulton_california_2021}] found to represent $\sim$~40\% of gas giants between 1 and 5 AU. Nevertheless, if we assume there are 5 undiscovered planets orbiting the ExEP targets between 1 and 5 EEID, we should expect approximately 2 of them to lie in their circumstellar HZs.}

\rev{There are caveats to how applicable our results are to future HWO observations.} We made two simplifying assumptions as part of our analysis: we placed every planet at quadrature phase (phase angle of $\pi/2$), and we assumed no OWA. We will first address the assumption of quadrature phase: [\citenum{savransky_parameter_2011}] show that, for a randomly oriented Keplerian orbit, phase angle $\beta$ follows a probability density distribution of $\sin{\beta}$ for $0\leq\beta\leq\pi$. We can estimate the mean value of the Lambertian phase function $\Phi(\beta)$ using this probability density; doing so yields a mean value $\overline{\Phi(\beta)}$ 0.56. This is larger than $\Phi(\pi/2)=1/\pi$, suggesting that our assumption of quadrature phase is in fact somewhat conservative. For reference, at $\Phi(\beta)=$~0.56, a planet on a 5 AU orbit 10 pc away would a projected separation of 0.45", well outside the presumed IWA of HWO's coronagraph \cite{stark_paths_2024}, so this mean value of \rev{$\Phi(\beta)$} should be representative of what HWO would actually be capable of observing.

We have further assumed that none of the gas giants in our simulations fall outside the OWA. Our analysis has shown that the ExEP target systems where broadband gas giant S/N, molecular detection S/N, and exomoon detection S/N are highest are those nearest to the Solar System. For a given projected physical separation, these same planets will have the largest working angles, and are therefore the most likely to fall outside the dark hole during a given observation. For HZ Jupiters, whose physical separations are by definition similar to those of exo-Earths, the fact that exo-Earth detection drives HWO's design ensures that they are unlikely to fall outside the OWA. However, a Jupiter twin at quadrature will have a working angle of 0.5” at 10 pc. The contrast of the simulated coronagraph used in [\citenum{stark_paths_2024}] exceeds $10^-9$ at $\sim$~28~$\lambda/D$, which for the observatory parameters adopted in our study ($\lambda_{cen}=$~\rev{550nm}, $D_{eff}=$~7.87~m) corresponds to an angular separation of 0.48”. The maximum fraction of one orbit that a planet with semi-major axis $a_p$ spends at a projected separation less than $x$ is given by $2\arcsin(x/a_p)/\pi=2\arcsin(\mbox{OWA}(a_p/d)^{-1})/\pi$, which to first order is directly proportional to the OWA. Given that [\citenum{fulton_california_2021}] found that the gas giant occurrence rate distribution peaks at approximately 3.6 AU, it is therefore crucial that the design of HWO’s coronagraphic instrument proceed with a desired yield in mind for gas giant exoplanet detections, molecular detections in their atmospheres, and exomoon detections.

\subsection{Exomoons and variability}
The high S/N that can be achieved for gas giants with HWO could also enable precise photometric time series observations of their atmospheres. This could be a unique science case for HWO as stable and accurate photometry of high-contrast companions is challenging from the ground (e.g. ELT).

In this study, we explore HWO's potential to detect exomoons around gas giant planets by comparing the achievable S/N of different detection methods. 
We showed that primary and secondary eclipses---\rev{where an exomoon and exoplanet cast shadows on each other}---are likely the most promising way \rev{for HWO to detect exomoons}. \rev{The strength of their} photometric signal\rev{s are similar to those of} primary and secondary transits---\rev{where the disks of the moon and planet obscure each other from our perspective}--- but the \rev{orbits of exomoons need only be nearly aligned with the plane of its planet's orbit, rather than with our line of sight. If the planet and moon's orbits are so aligned, some amount of the primary and secondary eclipses will be visible to us, though the strength of the photometric signal will vary.} We found that\rev{, by observing a single primary eclipse, HWO could detect Earth}-size moons around \rev{HZ Jupiters orbiting the nearest ($d<10\,$pc) ExEP target stars; with four orbits' worth of photometry, HWO could do the same for HZ Jupiters orbiting any ExEP star. Nine orbits could allow HWO to detect Mars-sized moons of HZ Jupiters, and Earth-sized moons of Jupiter twins, out to 10 pc.}

\rev{The occurrence rate of large exomoons around gas giant exoplanets is unknown; the only examples we have are the satellites of the Solar System gas giants. We choose to define ``large" exomoons as those with mass ratios $q\equiv M_{\leftmoon}/M_p\geq10^{-5}$, where $M_{\leftmoon}$ is the mass of the moon and $M_p$ is the mass of the parent planet. All four of Jupiter's Galilean moons (Io, Europa, Ganymede, and Callisto) are above this mass threshold ($2\cdot10^{-5}<q<8\cdot10^{-5}$); Saturn has one such moon, Titan ($q=2.4\cdot10^{-4}$); Uranus has four (Ariel, Umbriel, Titania, and Oberon, $10^{-5}<q<4\cdot10^{-5}$); and Neptune has one, Triton ($q=2.1\cdot10^{-4}$). Triton is most likely a captured Kuiper belt object that significantly disrupted Neptune's original system of satellites\cite{cuk_constraints_2005}, so Neptune's moon system may not be representative of those of Neptune-mass exoplanets. The high obliquity of Uranus's rotational axis was likely caused by a collision with a similarly sized planet, implying that its system of moons may have formed from the debris of this collision\cite{slattery_giant_1992}. Though such circumstances may be rare among exoplanets, the fact that Uranus's moon system bears strong similarity with Jupiter and Saturn's suggests that Uranus's moons may have formed by a similar mechanism.}

\rev{Like those of Jupiter, Saturn, and Uranus, the large moons of most gas giant exoplanets likely formed during the final stages of their parent planets' formation, in a circumplanetary disk (CPD) of gas and dust\cite{korycansky_numerical_1991,canup_formation_2002}. CPDs are expected to contain a total mass on the order of $10^{-2}M_p$, where $M_p$ is the mass of the planet they orbit\cite{moraes_growth_nodate,cilibrasi_satellites_nodate}. A dust-to-gas ratio consistent with the interstellar medium ($10^{-2}$) implies a CPD dust mass of $10^{-4}M_p$, consistent with the estimated mass of PDS70 b’s dust disk \cite{benisty_circumplanetary_2021}; this solid material accretes together to form moons.}

\rev{The currently favored model of moon formation in CPDs is that of ``sequential" satellite formation\cite{canup_common_2006, cilibrasi_n-body_2021}, in which multiple generations of moons form, migrate inwards due to gravitational interactions with the disk gas, and are engulfed by the planet. Any moons that have not been engulfed when the CPD gas dissipates remain in orbit of the host planet, with further dynamical evolution governed largely by gravitational interactions with the host star, host planet, and other moons. While moons are forming in the CPD and transporting material to the central planet, the CPD is likely also accreting additional material from the protoplanetary disk (PPD)\cite{pollack_formation_1996}; the total mass of the moons that remain after the CPD dissipates is therefore not constrained by the initial mass of the CPD.}

\rev{The mass at which a moon's type I migration timescale becomes comparable to the timescale of its continued accretion tends to be a constant fraction of the host planet's mass, of the order $M_{\leftmoon}\approx5\cdot10^{-5}M_p$ \cite{canup_common_2006}. The masses of individual satellites therefore likely scale with the masses of their planets. Simulations of the CPDs of gas giants formed by core accretion\cite{cilibrasi_n-body_2021} and gravitational instability\cite{inderbitzi_formation_2020} both imply that an individual satellite's probability of having a mass $M_{\leftmoon}\geq10^{-5}M_p$ is around 10\%, and that this probability scales as roughly $(M_{\leftmoon}/M_p)^{-1}$. These simulations also find that on average only 1 to 1.6 moons above this mass threshold should remain when the CPD dissipates, or $\sim$~0.1 to 0.16 moons with $M_{\leftmoon}\geq10^{-4}M_p$. By contrast, Jupiter, Saturn, and Uranus collectively have 9 of the former and 1 of the latter, suggesting an occurrence rate two to three times higher than the aforementioned simulations. Both [\citenum{cilibrasi_n-body_2021}] and [\citenum{inderbitzi_formation_2020}] average their occurrence rates over many simulations with widely varying dust-to-gas ratios and CPD refilling timescales from the PPD; [\citenum{cilibrasi_n-body_2021}] further find that only a small subset of this parameter space reproduces satellite systems similar to Jupiter's. Even so, the early CPDs of the Solar System gas giants may not be representative of CPDs in general.}

\rev{If we assume that the average mass of the 23 gas giants expected to orbit the ExEP stars is 1~$M_J$ (roughly in line with the median planet mass found by [\citenum{fulton_california_2021}]), the occurrence rates of [\citenum{cilibrasi_n-body_2021, inderbitzi_formation_2020}] imply that the ExEP gas giants are collectively orbited by approximately 4 moons more massive than Ganymede. If gas giant exoplanets form like those in our Solar System, we can multiply this number by 3 to suggest that this number might be closer to 12. As shown in Figure~\ref{fig:minimum_radius}, for a Ganymede-radius moon to be detectable with HWO it would need to orbit an HZ Jupiter that is itself less than $\sim$~7 pc from the Solar System. There are no known gas giants orbiting ExEP stars within this distance that have Earth-like instellations; because the nearest stars to Earth have been studied so closely\cite{harada_spores-hwo_2024}, it is unlikely that any more such gas giants will be found.}

\rev{With that said, the 7 pc detection limit from Figure~\ref{fig:minimum_radius} corresponds to moons with radii comparable to Ganymede; meanwhile, our occurrence rates are parameterized as a function of $q$. HWO can detect larger moons farther away, and such moons are most likely to exist around the largest planets. The minimum radius of a detectable exomoon is also smaller for a planet detected at a higher S/N, which will tend to favor gas giants orbiting more luminous stars. A reliable estimate of the exomoon ``yield" that HWO could achieve would require a yield study similar to those undertaken for exo-Earths \cite{kopparapu_exoplanet_2018,morgan_faster_2021,howe_scientific_2024, plavchan_analytic_2024, stark_paths_2024}, where HWO observations are simulated for exoplanets and exomoons randomly generated from joint mass, orbital radius, and orbital inclination distributions. For the purposes of this work, we chose to simplify the problem by focusing on gas giant exoplanets already known to orbit stars within 30 pc of the Solar System, attempting to rank them by how likely they are to host exomoons that HWO could detect.}
% \rev{information about gas giant planets already known to orbit nearby stars is available as part of the NASA Exoplanet Archive (NEA)\cite{christiansen_nasa_2025}. The host star properties used by EXOSIMS to simulate direct imaging observations is also available in the SIMBAD database\cite{wenger_simbad_2000}; with this information, we can simulate HWO observations of these gas giants and estimate their individual minimum exomoon radii and yields.}

\rev{We started by retrieving the masses (or $M\sin i$ values) and semi-major axes of all confirmed gas giant planets ($M$ or $M\sin i \geq0.1M_J$) at distances less than 30 pc from the Sun from the NASA Exoplanet Archive (NEA)\cite{christiansen_nasa_2025}. We used the published radii of these planets where available, and otherwise estimated their radii using the mass-radius relation for cold gas giants from [\citenum{thorngren_empirical_2019}]. We assumed that the albedos of these gas giants were determined solely by their effective temperatures, and assumed that our model albedo spectra of Jupiter twins and HZ Jupiters (see Figure~\ref{fig:mol_templates}) were representative of 150~K and 250~K gas giants, respectively. For higher-temperature gas giants ($T_{eff}=[300, 1000, 1450]$~K), we based our assumed albedos on the models of [\citenum{sudarsky_albedo_2000}] ($A_g=[0.1, 0.03, 0.55]$ respectively), which are in broad agreement with the optical albedos of observed hot Jupiters.\cite{esteves_changing_2015, schwartz_balancing_2015} If a $T_{eff}$ measurement was available in the NEA, we linearly interpolated between the above temperature-albedo pairs, and otherwise used each planet's instellation to find a $T_{eff}$ - albedo combination that was both consistent with our interpolation scheme and would bring the planet's blackbody radiation in equilibrium with its absorbed instellation (we assumed the planets had similar geometric and Bond albedos for simplicity).}

\rev{Based on these planetary radii and albedos, as well as parameters of the host stars taken from SIMBAD\cite{wenger_simbad_2000}, we simulated 5-minute HWO observations of these planets at quadrature phase and in the same bandpass as in Section~\ref{sec:detection}, giving us their detection S/N in each point in the light curve. We then used Equation~\ref{eq:minrad} to calculate minimum detectable $R_{\leftmoon}$ values for Galilean-like satellites around each planet, assuming each moon had an albedo of 0.2 and that each planet received 400 hours of photometric monitoring at 5-minute cadence. The simulations of [\citenum{canup_common_2006}] suggest that the semi-major axes of the Galilean moons are typical for large moons formed in CPDs; they also find that, for planets with different masses, their satellite semi-major axis distributions should scale with the planet's Hill radius $R_H = a_p\left( \frac{M_p}{M_*} \right)^{1/3}$, where $a_p$ is the planet's semi-major axis, $M_p$ is the mass of the planet, and $M_*$ is the mass of the host star. We therefore scaled the orbital radii of the Galilean moons by $(\frac{a_p}{5.2 \text{ AU}})(\frac{M_p/M_*}{M_J/M_{\odot}})^{1/3}$ to calculate the semi-major axes of each planet's four moons. Any moons that orbited at less than five times their host planet's radius were removed, assumed to have been engulfed by their parent planet when it was still cooling and contracting.}

\rev{To rank these planets according to their suitability for an exomoon search,}
% \rev{it is not enough to prioritize the planets with the lowest detectable exomoon radii. The large spherical moons of the gas giants in our solar system all exhibit similar moon-to-planet mass ratios\cite{canup_common_2006}; more massive planets should therefore be more likely to host larger moons. We therefore chose to roughly estimate an exomoon ``yield" for each known gas giant; the properties of the 10 planets with the highest yields are shown in Table~\ref{tab:moonhosts}. To estimate each planet's exomoon yield,}
\rev{we used the mass-radius relationship for terrestrial bodies from [\citenum{chen_probabilistic_2017}] to translate each minimum exomoon radius to a minimum exomoon mass. Then, in accordance with [\citenum{cilibrasi_n-body_2021, inderbitzi_formation_2020}], we assumed a conservative average number of moons per planet (i.e. a ``yield”) above some mass ratio threshold $q$ to be}
\begin{equation} \label{eq:yield}
    N(M_{\leftmoon}/M_p\geq q) = 
    \begin{cases}
        \dfrac{10^{-5}}{q} & \text{if } q\leq6\cdot10^{-3} \\
        0 & \text{otherwise.}
    \end{cases}
\end{equation}
\rev{Because we calculated the yields for four possible moons, we calculated the total conservative yield estimate by multiplying each individual yield by 0.25 and adding them together. For a more liberal estimate, we shifted the distribution forward in mass ratio by 0.5 dex. The names and relevant characteristics of the planetary systems with an expected total yield greater than $10^{-2}$ are shown in Table~\ref{tab:moonhosts}. Also listed are the minimum detectable exomoon radii for orbits analogous to Io and Callisto's, as well as our conservative and liberal exomoon yield estimates.}

\begin{table}[h!]
\centering
\caption{Properties of the known gas giant planetary systems in which HWO is most likely to photometrically detect an exomoon. Planets are ordered in ascending distance from the Sun (second column), and planets orbiting the same star are ordered in ascending semi-major axis. The instellations of each planet are shown in the fourth column; for planets in their stars' habitable zones, their instellations are displayed in bold green text. The minimum detectable radii of exomoons listed in the sixth column assume the parent planets are monitored for 400 hours, and that the moon has an albedo of 0.2. The expected exomoon yield assumes the satellite mass ratio distribution from [\citenum{cilibrasi_n-body_2021, inderbitzi_formation_2020}] and the mass-radius relation from [\citenum{chen_probabilistic_2017}].}
    \begin{tabular}{|p{3.2cm}|p{1.4cm}|p{1.4cm}|p{1.8cm}|p{0.85cm}|p{1.8cm}|p{1.9cm}|}
        
        \hline
        Name & Distance (pc) & Working angle (mas) & Instellation ($S_{\oplus}$) & Mass ($M_J$) & Minimum $R_{\leftmoon}$ ($R_{\oplus}$) & Expected moon yield ($10^{-2}$) \\
        \hline
        
        % \hline
        $\upsilon$ Andromedae Ac & 13.4 & 62 & 4.9 & 14.0 & 0.86 - 0.98 & 3.1 - 9.4 \\
        
        $\upsilon$ Andromedae Ad & 13.4 & 187 & \textcolor{OliveGreen}{\textbf{0.5}} & 10.2 & 0.78 - 1.13 & 4.6 - 13.7 \\
        
        % \hline
        47 Ursae Majoris b & 13.8 & 152 & \textcolor{OliveGreen}{\textbf{0.36}} & 2.5 & 0.78 - 1.13 & 0.7 - 2.0 \\
        
        % \hline
        % 47 Ursae Majoris c & 13.8 & 261 & 0.12 & 0.5 & 0.98 - 1.41 & 0.0 - 0.2 \\
        
        % \hline
        $\gamma$ Cephei Ab & 14.1 & 151 & 2.6 & 6.6 & 0.72 - 0.93 & 3.5 - 10.4 \\
        
        % \hline
        $\mu$ Arae e & 15.6 & 62 & 2.1 & 7.0 & 1.20 - 1.37 & 0.5 - 1.4 \\
        
        % \hline
        $\mu$ Arae b & 15.6 & 96 & \textcolor{OliveGreen}{\textbf{0.8}} & 4.3 & 0.93 - 1.06 & 0.7 - 2.1 \\
        
        % \hline
        $\mu$ Arae c & 15.6 & 276 & 0.11 & 4.4 & 1.31 - 1.89 & 0.5 - 1.4 \\
        
        % \hline
        % $\iota$ Horologii b & 17.3 & 53 & 2.1 & 6.2 & 0.95 & 0.03 \\
        
        % \hline
        % $\pi$ Mensae b & 18.3 & 180 & 0.13 & 14.1 & 1.79 - 2.57 & 0 - 1.0 \\
        
        % \hline
        % $\beta$ Pictoris c & 19.7 & 136 & 1.25 & 10.1 & 1.12 & 0.03 \\
        
        % \hline
        % $\epsilon$ Reticuli Ab & 18.3 & 70 & 4.39 & 1.6 & 1.04 & 0.006 \\
        
        % \hline
        % HD 33564 b & 21.0 & 52 & 2.83 & 9.1 & 1.01 & 0.04 \\
        
        % \hline
        % $\nu$ Octantis Ab & 21.2 & 58 & 8.6 & 2.2 & 0.39 - 0.8 & 0.4 - 1.1 \\
        \hline
\end{tabular}
\label{tab:moonhosts}
\end{table}

\rev{We stress that the yields in Table~\ref{tab:moonhosts} are intended only as heuristics for ranking planets according to HWO's approximate sensitivity to exomoon primary and secondary eclipses. These yields are especially sensitive to the assumed albedos of the planets, all of which are unknown. We also assumed that every moon's orbit was edge-on to our line of sight, though we did not incorporate primary and secondary transits when calculating the minimum $R_{\leftmoon}$. We also attempted to balance this assumption by assuming every gas giant was observed at quadrature phase, when in fact the observations could be timed to optimize the scattering angle of the planet and maximize its detection S/N in each exposure.}

\rev{By observing the four planetary systems in Table~\ref{tab:moonhosts} for 400 hours each with HWO, one could reasonably expect an exomoon yield of between 0.14 and 0.40; the minimum time investment required for this would be 1600 hours, or 67 days. However, because of the steep scaling of our yield estimates with minimum radius ($N\sim M_{\leftmoon}^{-1}\sim R_{\leftmoon}^{-3.6}$), the expected yield from an individual planet scales as roughly $t_{int}^{0.9}$. It may therefore make more sense to concentrate allocated exposure time on a single promising planetary system. The planets orbiting $\upsilon$ Andromedae A appear ideal for this purpose; planets d and c are the two planets we find to have the highest expected yields, and their photometric data could be gathered simultaneously if they both fall within the dark hole at the time of observation. Put together, their expected yield could be as high as 0.23 if their moons formed in CPDs similar to the Solar System gas giants. Furthermore, $\upsilon$ Andromedae Ad orbits in its host star's HZ; any large exomoons in orbit of this planet might therefore be similar to Earth, and would be ideal targets for spectroscopic studies of the kind discussed in Section~\ref{sec:moonspec}.}

\rev{However, the $\upsilon$ Andromedae A system is also where our assumption of edge-on orbits is most likely to break down; the planets' orbits are almost face-on to our line of sight, with $\upsilon$ Andromedae Ac and d having inclinations of approximately 8$^\circ$ and 24$^\circ$ respectively\cite{mcarthur_new_2010}. We account for this by assuming that the area of the moon's shadow is reduced by $\sin i$, and therefore we multiply each simulated moon’s primary eclipse strength by $\sin{i}$ to compensate. If $\upsilon$ Andromedae Ac and Ad possess coplanar moon systems, their combined exomoon yield after 400 h drops to 0.116. For their combined expected yield to reach 1, they would need to be observed for approximately 4,400 hours, or 180 days. While this would represent a significant time investment for HWO, this result demonstrates that HWO can constrain the presence of exomoon systems analogous to those in our Solar System orbiting nearby temperate gas giants on month-to-year timescales. Such observations could provide a new avenue to probe the dynamical histories of planetary systems, the properties of gas giant CPDs, and may lead to the discovery of large exomoons that are themselves deserving of characterization.}

\rev{Absent from our list are the planets in the $\beta$ Pictoris system, despite the fact that it is only 20 pc away and hosts three gas giant planets\cite{lagrange_unveiling_2020, gibbs_discovery_2026, sutlieff_direct_2026}, one of which is in the star's HZ. We omitted this planetary system because it is very young; there still exists a significant debris disk around it, which would add noise to an HWO light curve. The planets themselves are also self-luminous, meaning that their albedos do not accurately capture their brightnesses or detection S/N. Their thermal emission would also not be significantly changed by the passage of an exomoon's shadow over their surface, so the primary eclipse strengths we use in our calculations would not be accurate. However, the inclination of this system with respect to our line of sight is very high , consistent with 90$^{\circ}$ (i.e. edge-on). The probability that we could observe a primary and/or secondary exomoon transit is therefore much higher than in a system with an unknown, random inclination, and such an event would block the planet's own thermal emission, making this system another interesting target for exomoon searches.}
% \rev{Unlike for other, more mature gas giants, the suitability of each planet in the $\beta$ Pictoris system has little to do with the amount of instellation it receives, and more to do with its own radius and luminosity.}

\subsection{Exomoon atmospheres} \label{sec:atmo_discuss}
We also explore the potential of looking for the distinct spectral features of exomoon atmospheres. Since oxygen is not expected in the atmospheres of gas giants, detecting oxygen in the spectrum of an habitable zone gas giant would constitute evidence for an orbiting exomoon with an atmosphere. For a fixed exposure time, we find that the additional photon noise of the Jupiter-sized planet only decreases the oxygen detection S/N by a factor $3$--$5$ compared to a typical exoEarth. This suggests that detecting biosignatures around the moon of a habitable zone gas giant within $10\,$pc is as challenging as detecting oxygen in the atmosphere of an exoEarth at $\sim25\,$pc. Other molecules could similarly be detectable in the atmospheres of exomoons from deep spectroscopic observations of gas giant planets. This could not only lead the detection of exomoons, but also enable the characterization of their atmospheres. 
Notably, Saturn’s largest moon, Titan, has an atmosphere with a greater surface pressure than Earth’s, and has surface liquid in the form of methane and ethane \cite{stofan_lakes_2007}.

In the case where molecules could be found in either the planet or the moon, it might be possible to rely on both the large spin-broadening of the planet ($v\sin(i)\sim10\,$km/s for Jupiter) and the radial velocity shift of the moon compared to the planet given sufficient spectral resolution. For example, \rev{Io orbits Jupiter at a tangential speed of 17.5 km/s; assuming Io's orbit was edge-on to a telescope's line of sight, a spectrograph would need a spectral resolution of $\sim$~9,000 for lines in Io's reflection spectrum to shift by more than one full-width at half-maximum (FWHM). At high S/N it is possible to detect sub-FWHM shifts, but multiple adjacent spectral lines blending together can rapidly degrade RV sensitivity when R is decreased.} It might be possible to \rev{observe gas giants at high spectral resolution} without a dedicated coronagraphic high-resolution spectrograph; for example, by combining the coronagraph and the UV spectrograph ($100$--$1000$~nm; $R=500$--$50,000$ \cite{santos_ultraviolet_2025}) with an optical fiber similar to ground-based instruments like \rev{Keck/KPIC\cite{delorme_keck_2021}, VLT/HiRISE\cite{vigan_first_2024}, and Subaru/REACH\cite{kasagi_unveiling_2025}.}

The lessons learned from exploring the sensitivity of HWO to exomoons can be generalized to atmospheric variability studies \cite{sutlieff_exploring_2024} and even the search for rings around gas giants \cite{limbach_exomoons_2024}. From these simulations, we can infer that HWO could be sensitive to photometric variability at the 0.1-4\% level on one-hour timescales.
Sensitivity to $\sim$~1\% flux variations on 1 hour timescales could reveal rotation-induced surface variation comparable to that of Jupiter \cite{ge_rotational_2019}, allowing HWO to constrain the rotation rates of nearby gas giants. Better sensitivity could be achieved by phase folding multiple rotations of the planets (i.e., $\sim3$ sensitivity improvement in $\sim80\,$hours for Jupiter's 9-hour rotation period) and precisely constrain the surface features of their atmospheres.

\subsection{\rev{Priority of HWO Targets with Habitable Zone Gas Giants}} \label{sec:priority}

\rev{[\citenum{harada_spores-hwo_2024}] note that, if a sufficiently massive gas giant occupies an orbit in or near the HZ, the orbits of other planets in the HZ tend to be unstable; they suggest that stars known to host such gas giants should be removed from HWO's final target list or be given low observing priority. On the other hand, we have found in our simulations that these gas giants represent an opportunity for HWO to conduct novel and transformative scientific investigations, even with relatively small investments of observing time. In 10 hours, a R~=~1000 near-IR spectrograph on HWO could achieve $\langle$S/N$_i\rangle\gtrsim$~10 for HZ Jupiters (see Figure~\ref{fig:bin_snr_res_exp}), and could easily detect \methane{}, \water{}, and \ammonia{}. Placing constraints on the abundances of these molecules, \ammonia{} in particular, would allow the exoplanet community to test the predictions of planet formation theories with old, low-mass gas giants, a planet demographic that in many cases cannot be spectroscopically characterized with current instrumentation. HWO could also achieve photometric observations of HZ Jupiters at high cadence and precision, making HZ Jupiters prime targets for variability studies. This could further advance planet formation investigations by allowing the determination of gas giant spin rates and axial tilt, as well as potentially enabling the photometric detection of exomoons. Over the course of a deep ($t_{int}\approx$~400 hours) spectroscopic observation, HWO spectra of HZ Jupiters may have such high S/N that they enable the detection of absorption features in the atmosphere of an orbiting exomoon. For these reasons, we argue that HWO observations of gas giants orbiting in the vicinity of the HZ may be more scientifically rewarding than many studies have assumed.}

\section{Conclusion} \label{sec:conclusion}

Once launched in the \rev{2040s}, the Habitable Worlds Observatory will be one of the most powerful astronomical observing platforms in history. The ability to image Earth-like exoplanets less than $10^{-10}$ times as bright as their stars at working angles less than 0.1” will also allow HWO to revolutionize the study of the atmospheres and satellite systems of gas giants at a wide variety of instellations around nearby (\rev{$d\lesssim30$~pc}) stars. In this work, we use EXOSIMS \cite{savransky_wfirst-afta_2015} and custom Python code to explore the gas giant-specific science cases that would become feasible with high S/N HWO spectra and photometry. We find that:
\begin{enumerate}
    % \item At quadrature phase, gas giant exoplanets with Jupiter-like radii, albedos, and instellation have planet-to-star flux ratios between $2\cdot10^{-10}$ and $3.5\cdot10^{-8}$. HWO would receive comparable flux from these planets as from the brightest exozodi in our simulation, and are 2 - 600 times as bright as our simulated Solar System zodiacal light. Jupiter-like planets at 1 EEID are $\sim$~20 times brighter than similar planets at 5 EEID.

    \item \rev{To detect the reflected light of Earth-like planets orbiting any ExEP target star in 10 hours or less (assuming a 20\% bandpass centered at 550nm and a primary mirror diameter of 7.87m), HWO will need to simultaneously achieve raw coronagraphic starlight suppression of $10^{-10}$ and speckle PSF stability at the 1\% level. For Jupiter-sized planets orbiting at 5 EEID from ExEP stars, these same observations would result in detection S/N values between 90 and 1000. For Jupiter-sized planets orbiting at 1 EEID, S/N values would range between 300 and 10,000.}
    %For both types of gas giant, increasing integration time to 10 and 400 hours multiplies the detection S/N by a factor of 10 and 60, respectively.
    
    \item In 10-hour integrations with an R~=~100 spectrograph, HWO will be able to detect \rev{atmospheric \ammonia{} in the atmosphere of a Jupiter twin orbiting any target star with S/N~$\gtrsim$~15. For an instrument with R~=~10,000 and an EMCCD detector with dark current of $3\cdot10^{-5}\text{e}^-\text{pix}^{-1}\text{s}^{-1}$, HWO should be able to detect \ammonia{} at S/N~$\gtrsim$~5.}
    % in the atmospheres of Jupiters at 5 EEID  if their host stars are $<$~13 pc from the Sun. For Jupiters at 1 EEID, ammonia is detectable at S/N~$>$~30 for all simulated planets.  Increasing the instrumental spectral resolution from 100 to 1000 multiplies the \ammonia{} detection S/N by a factor of 1.4. Higher-resolution model spectra of gas giant atmospheres would allow us to explore the trade space of optimal spectral resolution.
    
    \item \rev{Time series} broadband flux measurements of gas giants in reflected light could allow HWO to investigate gas giant time variability. At \rev{550nm}, HWO is sensitive to $\leq$~1\% \rev{flux} changes in 1-hour integrations for all simulated HZ Jupiters, and for Jupiter twins within \rev{7} pc of the Solar System.
    
    % \item At a central wavelength of \rev{550nm} and a spectral bandwidth of 20\%, simulated Jupiter-like exoplanets orbiting at 5 EEID have broadband detection S/N between 8 and \rev{150} after 5 minutes of integration time. At 1 EEID, detection S/N ranges from 30 to 700.
    
    \item \rev{Photometric time series at a cadence of $\sim$~5 minutes could allow HWO to detect single primary eclipses of Earth-sized exomoons with short ($\sim$~42 hour) orbital periods around HZ Jupiters within roughly 10 pc. With four orbital periods' worth of observations, HWO could detect Earth-sized exomoons orbiting HZ Jupiters around any ExEP star; nine orbits could allow the detection of Mars-sized moons within 10 pc.}
    % exomoons are detectable out to a distance of $>$~25\rev{, and for Ganymede-like exomoons out to} 8 parsecs.
    
    \item Assuming that the \rev{spectrum of an HZ Jupiter} can be near-perfectly modeled in the 675-825~nm region, oxygen could be detected in the atmosphere of an Earth-sized exomoon out to a distance of \rev{8 pc, even at a spectral resolution of 10,000. Such high spectral resolution would help break degeneracies between exomoon atmospheric signatures and alternative explanations for the oxygen signature; a high-resolution spectrograph on HWO might therefore enable it to search for exomoons similar to Earth.}  
\end{enumerate}
To conclude, deep spectroscopic observations of gas giant planets in the solar neighborhood with HWO could yield unprecedented insight about the atmospheres of gas giant planets, their formation and even their moons.

%\section{Appendix} \label{sec:appendix}

% \disclosures 
\subsection*{Disclosures}
The authors declare that there are no financial interests, commercial affiliations, or other potential conflicts of interest that could have influenced the objectivity of this research or the writing of this paper. Minor use of ChatGPT was made to assist in coding, figures, and writing latex equations.

\subsection* {Code, Data, and Materials Availability} 
An example Python script is provided on Github\footnote{\url{https://github.com/jruffio/speckles_snr_gain}}.

\subsection* {Acknowledgments}
 Material presented in this work is supported by the National Aeronautics and Space Administration under Grants/Contracts/Agreements No. 80NSSC25K7300 (B.D., J.-B.R.) issued through the Astrophysics Division of the Science Mission Directorate. Any opinions, findings, and conclusions or recommendations expressed in this work are those of the author(s) and do not necessarily reflect the views of the National Aeronautics and Space Administration. \rev{This research has made use of the NASA Exoplanet Archive, which is operated by the California Institute of Technology, under contract with the National Aeronautics and Space Administration under the Exoplanet Exploration Program. This research has made use of the SIMBAD database,
operated at CDS, Strasbourg, France.} S.S. acknowledges support from an STScI Postdoctoral Fellowship. B.D. would like to thank Charlie Goldberg for discussion icy moon surface albedos and spectral features.

%%%%% References %%%%%

\bibliography{report}   % bibliography data in report.bib
\bibliographystyle{spiejour}   % makes bibtex use spiejour.bst

\listoffigures
% \listoftables

\end{spacing}
\end{document}